\documentclass[reprint,
amsmath, amssymb, onecolumn, preprintnumbers,
aps, nofootinbib]{revtex4-2}

\usepackage[utf8]{inputenc}
\usepackage{epsf, epsfig, subfigure, graphicx}
\usepackage{amsfonts, amsmath, amssymb}
\usepackage{latexsym}
\usepackage{color, xcolor}
\usepackage{accents}
\usepackage{dhucs}
\usepackage{hyperref}
\usepackage{verbatim}
\usepackage{slashed}
\usepackage{notes2bib}
\usepackage{tikz}
\usepackage{mathtools}
\usepackage{float}
\usepackage{empheq}

\def\bea{\begin{eqnarray}}
\def\eea{\end{eqnarray}}

\def\GeV{\,{\rm GeV}}

\def\eV{\,{\rm eV}}

\def\eps{\epsilon}

\makeatletter 
    
\renewcommand\onecolumngrid{
\do@columngrid{one}{\@ne}%
\def\set@footnotewidth{\onecolumngrid}
\def\footnoterule{\kern-6pt\hrule width 1.5in\kern6pt}%
}

\renewcommand\twocolumngrid{
        \def\footnoterule{
        \dimen@\skip\footins\divide\dimen@\thr@@
        \kern-\dimen@\hrule width.5in\kern\dimen@}
        \do@columngrid{mlt}{\tw@}
}%

\makeatother    

\begin{document}

\title{Gravitational Interaction of Ultralight Dark Matter with Interferometers}
\author{Hyungjin Kim}
\email{hyungjin.kim@desy.de}
\affiliation{Deutsches Elektronen-Synchrotron DESY, Notkestr. 85, 22607 Hamburg, Germany}
\preprint{DESY-23-085}

\begin{abstract}
Ultralight dark matter exhibits an order-one density fluctuation over the spatial scale of its wavelength. These fluctuations gravitationally interact with gravitational wave interferometers, leading to distinctive signals in detectors. We investigate the ultralight dark matter-induced effects in the gravitational wave interferometers. We perform a systematic computation of the power spectrum of ultralight dark matter in interferometers. We show that the ultralight dark matter-induced effect is most relevant for the interferometers with long baseline and that it is only a sub-leading effect compared to the estimated noise level in the case of Laser Interferometer Space Antenna or future interferometers with an arm-length comparable to a few astronomical units. Gravitational wave interferometers can then place upper limits on the ultralight dark matter density in the solar system. We find that, under certain assumptions, future interferometers with AU-scale arm-length might probe the dark matter density a few hundred times the local dark matter density, which is measured over a much larger spatial scale. 
\end{abstract}

\maketitle
\onecolumngrid
\tableofcontents

\section{Introduction}
Ultralight dark matter (ULDM) remains an attractive dark matter candidate.
Its mass is smaller than several eV-scale, and it behaves like classical waves. 
The QCD axion is one of the most theoretically interesting ultralight dark matter candidates~\cite{Peccei:1977hh, Weinberg:1977ma, Wilczek:1977pj, Preskill:1982cy, Abbott:1982af, Dine:1982ah}.
Other interesting candidates often arise from recent developments in dynamical solutions to the electroweak hierarchy problem~\cite{Graham:2015cka, Arvanitaki:2016xds, Banerjee:2018xmn, Banerjee:2020kww, Arkani-Hamed:2020yna, TitoDAgnolo:2021nhd, TitoDAgnolo:2021pjo, Chatrchyan:2022dpy}. Phenomenologically, ultralight dark matter of extremely small mass $m\sim 10^{-21}\eV$ has also been actively investigated as it changes the prediction of cold dark matter at galactic and sub-galactic scales~\cite{Hu:2000ke}.
Being wave-like, such ultralight dark matter candidates lead to interesting phenomenology in the early and late universe. See Ref.~\cite{Hui:2021tkt} for a recent review on wave dark matter.

Recent numerical simulations of ultralight dark matter halo have observed an order one density fluctuation all over galaxies~\cite{Schive:2014dra, Schive:2014hza}.
The fluctuation is extended over the spatial scale given by the wavelength of dark matter. 
This characteristic density fluctuation can be intuitively understood in terms of {\it quasiparticles}, whose size ($\lambda$) and mass $(m_{\rm eff})$ are given as its wavelength and the total mass enclosed within the volume of de Broglie wavelength~\cite{Hui:2016ltb}:
\begin{align}
\lambda &= \frac{1}{m \sigma} \simeq 10^2\,{\rm pc} \Big(\frac{10^{-22}\eV}{m}\Big)
\Big( \frac{200\,{\rm km/sec}}{ \sigma} \Big) , 
\\
m_{\rm eff} &= \frac{\pi^{3/2} \rho_0}{(m\sigma)^3} 
\simeq 5\times 10^4 M_\odot 
\Big( \frac{\rho_0}{{\rm GeV/cm^3}} \Big)
\Big( \frac{10^{-22}\eV}{m} \Big)^3
\Big( \frac{200\,{\rm km/sec}}{ \sigma} \Big)^3 .
\end{align}
Here $m$ is the mass of ULDM, $\rho_0$ is the mean dark matter density, and $\sigma$ is the velocity dispersion. 
For $m\ll \eV$, the size and the mass of the quasiparticles could be astronomical.\footnote{The quasiparticle, however, is not a particle in a traditional sense. The density fluctuation is a result of wave interference, and it lasts only for a coherence time $\tau = 1/m\sigma^2$. In addition, the fluctuation of the mass within a fixed volume in galaxies follows the exponential distribution in the case of ULDM, while it is the Poisson distribution for the particle dark matter. These features make the behavior of ULDM qualitatively different from that of massive particle dark matter candidates, such as primordial black holes or massive compact halo objects, which have been previously studied in~\cite{Seto:2004zu, Adams:2004pk, Hall:2016usm, Jaeckel:2020mqa, Baum:2022duc}.}

These quasiparticles interact continuously with interferometers, designed to detect gravitational waves (GWs) of astrophysical and cosmological origin. 
Since they are engineered to measure an extremely small disturbance of spacetime due to GWs, any interaction of quasiparticles with such interferometers leaves some distinctive impacts on detectors. The main goal of this work is, therefore, to characterize the impacts of ULDM on gravitational wave interferometers. In particular, we ask two questions: (i) if ULDM signals can be larger than the projected noise level in each detector and (ii) if the dark matter density in the solar system can be probed only through the gravitational interaction of order-one density fluctuations of ULDM with detectors. We will address these questions by computing the signal power spectrum of ULDM in detectors and comparing it with the noise power spectrum of detectors.

Without detailed computation, we can already estimate the expected effects of ULDM and partially answer one of the above two questions. 
Consider an interferometer with two test masses.
A useful quantity is a differential acceleration, $\Delta a = a_1 - a_2$, between two test masses induced by ULDM density fluctuations.
Suppose that a quasiparticle is located right next to the first test mass.
In this case, the differential acceleration is estimated as
$$
\Delta a = a_1 - a_2 = \frac{Gm_{\rm eff}}{\lambda^2} - \frac{Gm_{\rm eff}}{(L + \lambda)^2}
\simeq 
\bar a
 \min\Big(1 , \frac{2L}{\lambda} \Big) , 
$$
where $L$ is a typical size of the interferometer and typical acceleration $\bar a$ is defined as
\bea
\bar a \equiv \frac{Gm_{\rm eff}}{\lambda^2} = \pi^{3/2} G\rho_0 \lambda. 
\eea
As the mass $m$ decreases (hence wavelength and effective mass increases), the differential acceleration increases until it saturates to $\bar a = 2 G \rho_0 L$ when $\lambda =2L$; when the size of a quasiparticle becomes comparable to the size of the interferometer.
At this saturation point, we find $\Delta a \sim 10^{-28} \,{\rm m\, s^{-2}}$ for $L = 4 \,{\rm km}$ (LIGO/VIRGO), $\Delta a \sim 10^{-22}\,{\rm m\,s^{-2}}$ for $L = 2.5 {\rm M}\,{\rm km}$ (LISA), and $\Delta a \sim 10^{-20}$\ for $L = 400 {\rm M}\,{\rm km}$ (e.g. $\mu$Ares strawman mission concept~\cite{Sesana:2019vho} and Fedderke et al~\cite{Fedderke:2021kuy}). 
Here we use $\rho_0 = 0.4 \GeV/{\rm cm^3}$. 

This can be compared with the sensitivity of detectors.
The strain noise power spectrum can be converted into the root-mean-square fluctuation of acceleration by $\Delta a \sim [ S_n (2\pi f)^4 L^2 \Delta f]^{1/2}$. 
We find $\Delta a = 10^{-14}\,{\rm m\, s^{-2}}$ for LIGO with $f \sim \Delta f \sim 50\,{\rm Hz}$, $\Delta a = 10^{-16}\,{\rm m\, s^{-2}}$ for LISA with $f \sim \Delta f \sim 0.1{\rm mHz}$, and $\Delta a = 10^{-17}\,{\rm m\, s^{-2}}$ for $\mu$Ares proposal with $f \sim \Delta f \sim  \textrm{a few} \times 10^{-7}{\rm Hz}$.
For ground-based interferometers, the ULDM-induced noise is irrelevant as it is many orders of magnitude smaller than detector noises.
For space-borne interferometers, the ULDM effect is a few orders of magnitude smaller than the mission-required detector noise level.
We may conclude from this comparison that the ULDM effect in interferometers through gravitational interaction is unlikely to interfere with the operation of interferometers for the detection of GWs.
For a more correct comparison, however, we will need to compute the power spectrum of ultralight dark matter-induced effects since the above estimation on $\Delta a$ does not carry any spectral information about the ULDM effects in the detector.

\begin{figure}[t]
\centering
\includegraphics[width=0.45\textwidth]{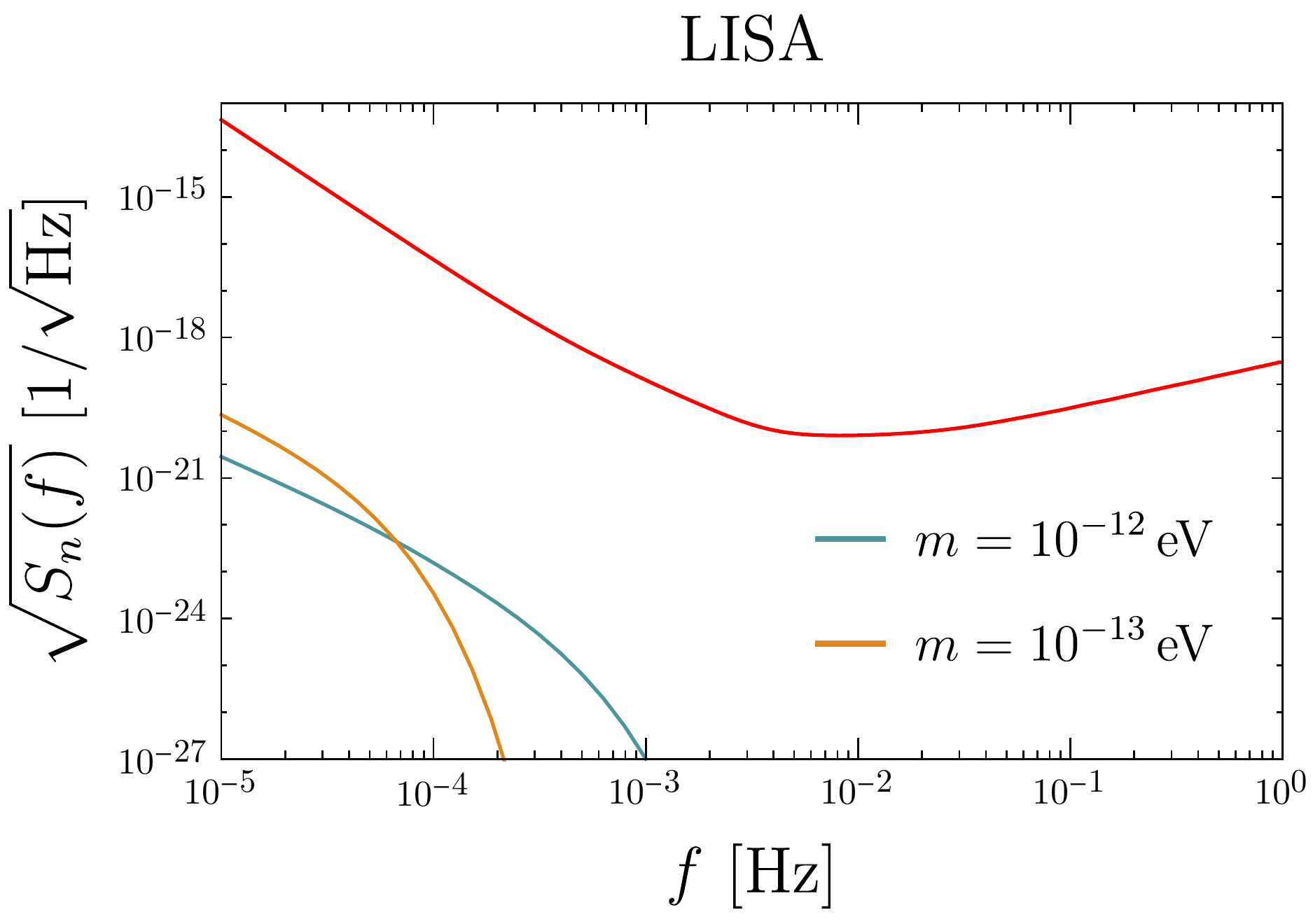}
\quad
\includegraphics[width=0.45\textwidth]{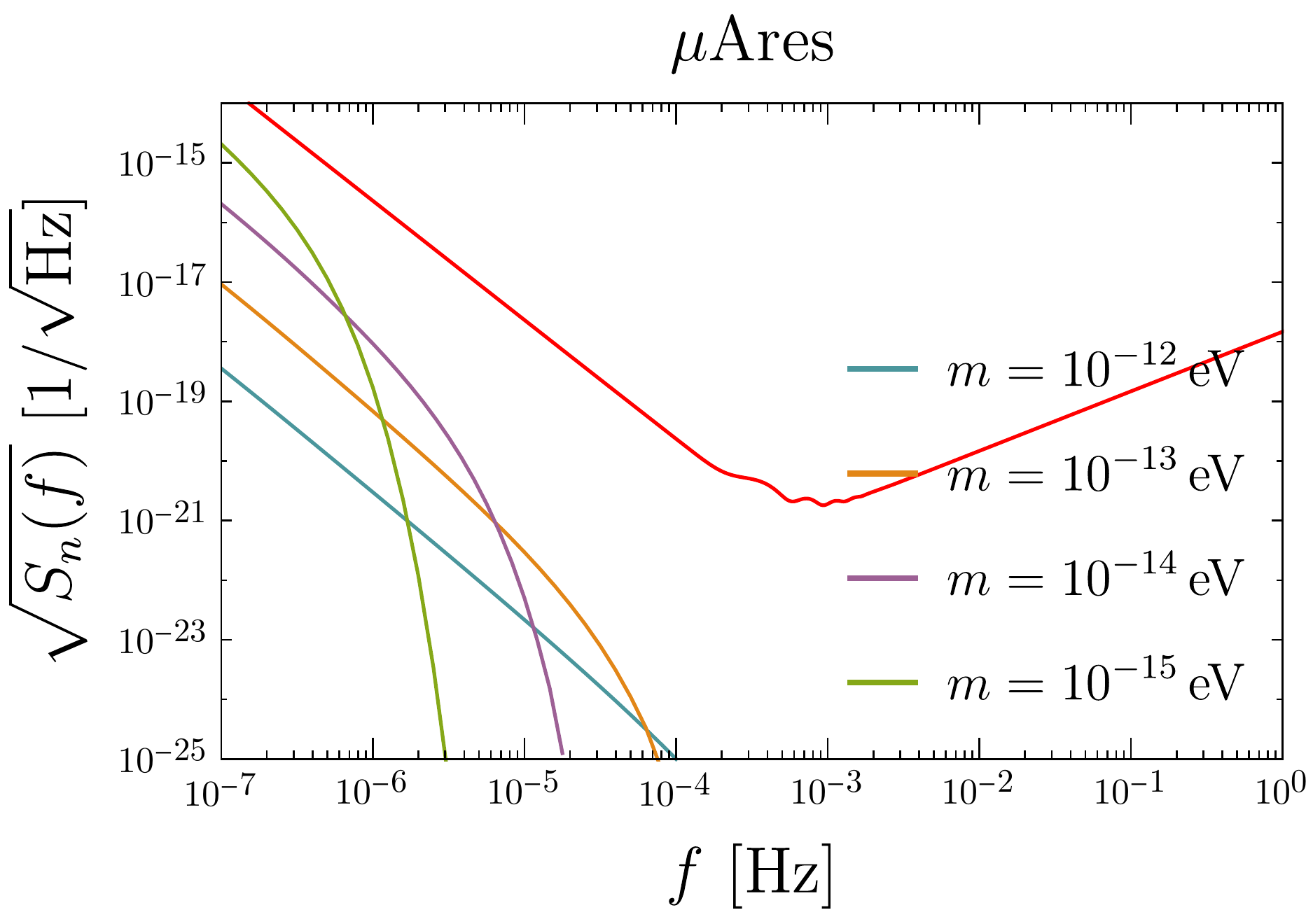}
\caption{The ultralight dark matter-induced noise for
(left) LISA and (right) $\mu$Ares proposal~\cite{Sesana:2019vho}. 
The red line is the required noise level for the mission. 
The other colored solid lines are the ultralight dark matter-induced noise for given masses. 
In both cases, the shown is the strain noise power spectrum in the Michelson-like TDI-X variable. 
Note that, for $\mu$Ares, the low-frequency noise could be dominated by the gravity gradient noise as discussed in Ref.~\cite{Fedderke:2020yfy}. 
See the main text for details.}
\label{fig:ULDM_noise}
\end{figure}

\begin{figure}
\centering
\includegraphics[width=0.6\textwidth]{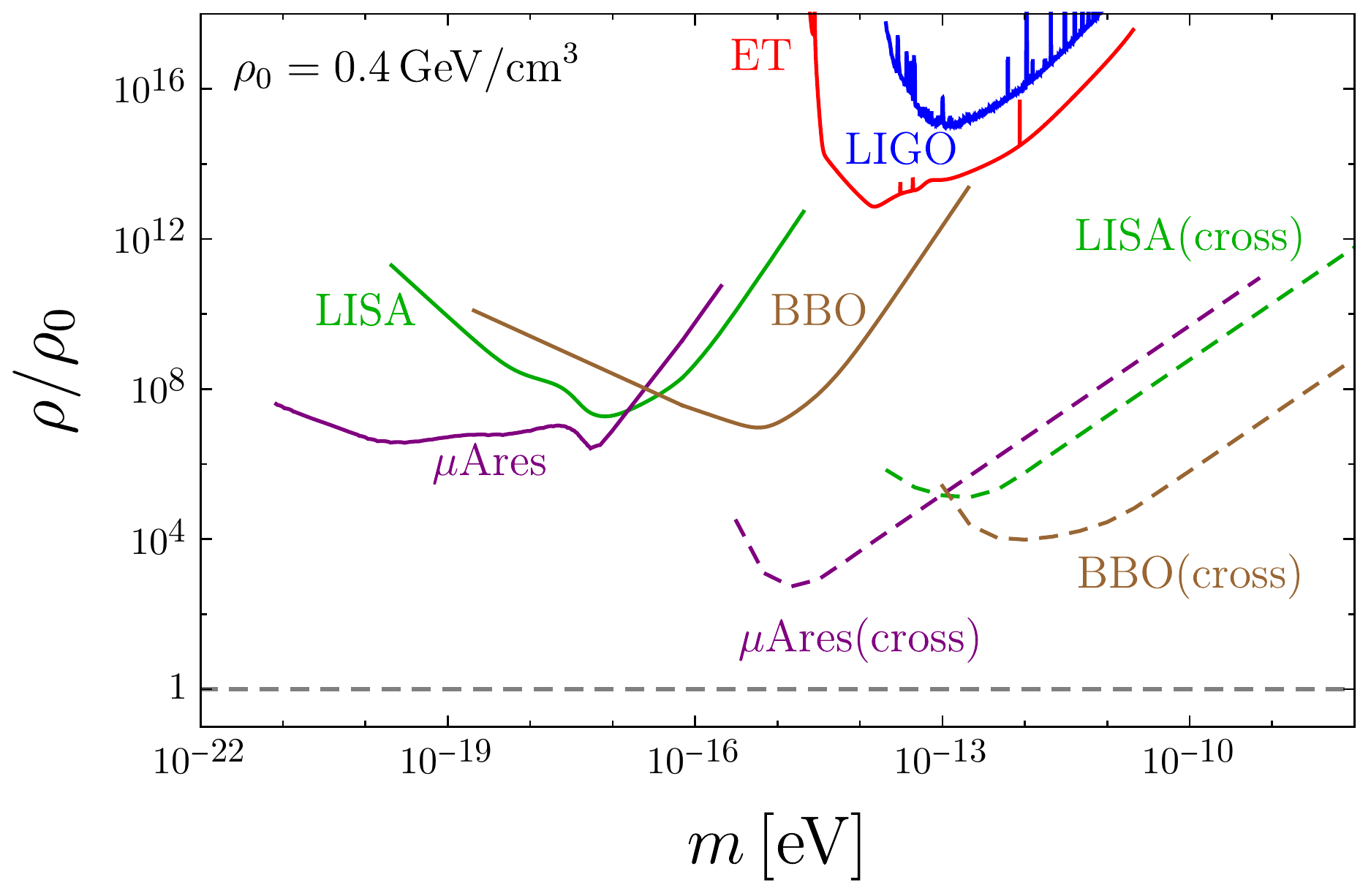}
\caption{The constraints and projections on dark matter density in the solar system. 
Here $\rho_0 = 0.4\GeV/{\rm cm^3}$ is the local dark matter density measured over $\gtrsim {\cal O}(100)$~pc scale (see e.g.~\cite{Read:2014qva, deSalas:2020hbh} for reviews). 
The solid lines are obtained from the matched filter for the coherently oscillating dark matter signal at $\omega =2m$, while the dashed lines are obtained by cross-correlating signals in two detectors at $\omega < m\sigma^2$. 
Note that the current design of LISA does not include two detectors, but other space-borne interferometer proposals, such as TianQin or Taiji, would potentially allow the cross-correlation searches~\cite{Cai:2023ywp}.
For the matched filter search of the coherently oscillating signal, we take for simplicity vanishing dark matter velocity $v_0=0$; we check this approximation does not change the numerical result in a significant way. 
For the stochastic signal search with the cross-correlation, we take an average of the signal over twelve months. 
See the main text for details.
}
\label{fig:summary}
\end{figure}

We summarize the main results here. 
From the detailed computation of the ULDM power spectrum, we find that the spectrum generally consists of two distinctive frequency components: one at $\omega = 2m$ and the other at $\omega < m \sigma^2$.
This has interesting phenomenological implications as it offers different ways to search for ultralight dark matter signals. 
In Figure~\ref{fig:ULDM_noise}, we show the low-frequency part ($\omega < m\sigma^2$) of the ULDM signal. 
The ULDM noise is sub-dominant even for the interferometers with arm-lengths larger than a few million km, such as LISA and other proposals with arm-length comparable to astronomical units, confirming our order-of-magnitude estimation above. 
The interferometers then can be used to place an upper limit on the dark matter content in the solar system.
Two distinct frequency components offer multiple ways for ULDM searches in interferometers.
The $\omega =2m$ component behaves similarly to a deterministic signal, which can be searched with the matched filter.
The low-frequency ($\omega < m\sigma^2$) component, on the other hand, behaves similarly to the stochastic background, which can be searched by cross-correlating detector outputs if more than two detectors are available.
The resulting constraints and projections on $\rho/\rho_0$ are shown in Figure~\ref{fig:summary}, suggesting that the density of ULDM in the solar system, $\rho \sim {\cal O}(10^2)\rho_0$, could be probed with GW interferometers only through gravitational interaction.

The work is organized as follows.
In section~\ref{sec:ULDM}, we review the basic statistical properties of ultralight dark matter and compute the spectrum of density contrast. 
This will establish a basis for the detailed investigation of the response of GW interferometers to the ULDM density fluctuations. 
In section~\ref{sec:GW_int}, we discuss how GW interferometers respond to the ULDM density fluctuation, and how the ultralight dark matter signal can be searched with the matched filter and with cross-correlation of detector outputs in cases where there are more than two detectors. 
In section~\ref{sec:discussion}, we discuss some of the assumptions that might affect our results. 
We conclude in section~\ref{sec:conclusion}. 
The natural unit ($c = \hbar =1$) is used in this work. 

\section{Ultralight dark matter}\label{sec:ULDM}
We review the basic statistical properties of ultralight dark matter in this section. 
We consider a scalar field minimally coupled to gravity without self-interaction. 
The ultralight dark matter field operator is given as
$$
\hat \phi (x) = 
\sum_i \frac{1}{\sqrt{2mV}} 
( a_i e^{-i k_i \cdot x} + a_i^\dagger e^{i k_i \cdot x} ) ,
$$
where $a_i$ and $a^\dagger_i$ are annihilation and creation operator of the plane wave mode $i$, satisfying $[a_i, a^\dagger_j] = \delta_{ij}$. 
The continuum expression can be obtained by $\sum_i \to V \int d^3k / (2\pi)^3$, $a_i \to a(\vec k) / \sqrt{V}$, $[a(\vec k), a^\dagger(\vec q)] = (2\pi)^3 \delta^{(3)}(\vec k - \vec q)$. 
For later convenience, we stick to the above discrete convention. 
Any deformation of wave function due to the gravitational potential of the Sun is ignored since this effect is at the level of ${\cal O}(1)\%$ for the halo dark matter~\cite{Kim:2021yyo}. 

The statistical property of ultralight dark matter is defined by the density operator.
The density operator is $\hat\rho = \prod_i \hat\rho_i \otimes$ where $\hat \rho_i$ is the density operator of each mode, given by~\cite{Kim:2021yyo}
\bea
\hat\rho_i = \int d^2\alpha_i \, p(\alpha_i)  | \alpha_i \rangle | \alpha_i |.
\label{den_op_single}
\eea
Here $|\alpha_i \rangle$ is the coherent state, satisfying $a_i |\alpha_i \rangle = \alpha_i | \alpha_i \rangle$ ($\alpha_i \in {\mathbb C}$), and the quasi-probability distribution $p(\alpha_i)$ is given as
\bea
p(\alpha_i) = \frac{1}{\pi f_i} \exp\left[ - \frac{|\alpha_i|^2}{f_i} \right].
\label{quasi_prob}
\eea
The $f_i$ is the mean occupation number for the mode $i$. 
This quasi-probability distribution function $p(\alpha_i)$ corresponds to the Rayleigh distribution for the amplitude, $|\alpha_i|$, and the uniform distribution for the phase, ${\rm arg}(\alpha_i)$~\cite{Derevianko:2016vpm, Foster:2017hbq, Centers:2019dyn}.
With this description, any $n$-point correlation function of an operator $\hat\phi$ can be easily computed by the trace operation, $\langle {\cal O} \rangle = {\rm Tr}({\cal O} \hat\rho)$. 
In the large occupation number limit, the operators $(a_i, a_i^\dagger)$ can be replaced by commuting complex random number $(\alpha_i, \alpha_i^*)$, whose probability distribution is given by \eqref{quasi_prob}.

It is useful to note the ensemble averages of the creation and annihilation operators:
\bea
\langle a_i a^\dagger_j \rangle = \langle a_i^\dagger a_j \rangle = f_i \delta_{ij}, 
\quad
\textrm{and}
\quad
\langle a_i a_j a_k^\dagger a_\ell^\dagger \rangle = f_i f_j
( \delta_{ik} \delta_{j\ell} + \delta_{i\ell} \delta_{jk} ) . 
\eea
Here we approximate $a_i$ and $a_i^\dagger$ as a commuting random complex number $\alpha_i$ and $\alpha_i^*$, whose underlying distribution follows the quasi-probability distribution $p(\alpha_i)$. 
The ultralight dark matter field defined with~\eqref{quasi_prob} is a Gaussian random field, and therefore, the ensemble average of any operators with $N(a) - N(a^\dagger) \neq 0$ vanishes, where $N(a)$ and $N(a^\dagger)$ are the number of annihilation and creation operator, respectively.

The energy density of the ultralight dark matter is
\begin{align}
\rho_\phi  &= \frac{1}{2}
\left[ \dot{\phi}^2 + (\nabla \phi)^2 +  m^2 \phi^2
\right]
\nonumber\\
&=
\frac{1}{2 V}
\sum_{i,j}
\frac{1}{ \sqrt{2 \omega_i} \sqrt{2\omega_j} }
\Big[
W_{ij}^- 
\Big( 
a_i a_j e^{-i ( k_i + k_j)\cdot x}
+ a^\dagger_i a^\dagger_j e^{i ( k_i + k_j)\cdot x}
\Big)
+ W_{ij}^+ 
\Big(
a_i a_{j}^\dagger e^{-i ( k_i - k_j )\cdot x}
+ a^\dagger_i a_{j} e^{i ( k_i - k_j )\cdot x}
\Big)
\Big]. 
\label{density_operator}
\end{align}
For notational simplicity, we have introduced $W^\pm_{ij}$, defined as
\begin{align}
W^\pm_{ij} = 
m^2 \pm ( \omega_{k_i} \omega_{k_j} + \vec k_i \cdot \vec k_j)
\approx
2m^2 \times
\begin{cases}
1 & \textrm{for $W_{ij}^+$}
\\
- \frac{1}{4} ( \vec v_i + \vec v_j)^2 & \textrm{for $W_{ij}^-$}
\end{cases}
\end{align}
where the second expression holds in the non-relativistic limit. 
The mean energy density in the non-relativistic limit is
\bea
\langle \rho_\phi \rangle \equiv  \rho_0
\approx \frac{m}{V} \sum_{i} f_i
= m \int \frac{d^3k}{(2\pi)^3} f(\vec k) 
= m \int d^3v f (\vec v), 
\label{normalization_f}
\eea
By taking the continuum limit, we reproduce the usual expression for the local dark matter density.
In the last line, we change the integration variable to the velocity. 
The above expression fixes the normalization of the mean occupation number $f(\vec v)$ used in this work.

\subsection{Density fluctuation}
Let us consider the density fluctuation, $\delta\rho_\phi=  \rho_\phi -  \rho_0$. 
The mean value vanishes $\langle \delta \rho_\phi \rangle = 0$.
The correlator is 
$$
\langle \delta\rho_\phi(x) \delta\rho_\phi(y) \rangle = \int \frac{d^4k }{ (2\pi)^4 }e^{-i k \cdot (x-y)} P_{\delta\rho}(k), 
$$
where the power spectrum is defined as $\langle \delta\rho_\phi(k) \delta\rho_\phi^*(k') \rangle = (2\pi)^4 \delta^{(4)} (k - k ') P_{\delta\rho}(k)$ and given by
\begin{align}
P_{\delta\rho}(\omega,\vec k) 
=&
\frac{(2\pi)^4}{2 }
\int d\Pi_1 d\Pi_2
f(\vec p_1) f(\vec p_2) 
\nonumber
\\
\times &  \Big[ 
(W^-_{12})^2 
\big(
 \delta^{(4)}(k - p_1 - p_2) 
+ \delta^{(4)}(k + p_1 + p_2)  
\big)
 + (W^+_{12})^2 
\big( 
\delta^{(4)}(k - p_1 + p_2) 
+ \delta^{(4)}(k + p_1 - p_2)  
\big)
\Big]. 
\label{deltarho_PS}
\end{align}
This can be straightforwardly computed by using \eqref{quasi_prob} and \eqref{density_operator}.
Here $d\Pi = [d^3p/(2\pi)^3] (2E)^{-1}$ is the Lorentz invariant phase space measure. 
In the frequency space, the correlator is
\bea
\langle 
\widetilde{\delta \rho}_\phi(\omega , \vec x) 
\widetilde{\delta \rho}_\phi^*(\omega', \vec y)
\rangle
= (2\pi) \delta(\omega - \omega') S_{\delta\rho}(\omega, \vec L) , 
\eea
where $\vec L = \vec x - \vec y$ and the power spectrum $S_{\delta\rho} (\omega, \vec L)$ is given by
\bea
S_{\delta\rho}(\omega, \vec L) 
= \int \frac{d^3k}{(2\pi)^3} e^{i \vec k \cdot \vec L} 
P_{\delta\rho}(\omega, \vec k). 
\label{S_P}
\eea
This power spectrum $S_{\delta\rho}(\omega,\vec L)$ in the frequency space turns out to be more useful for the study of the response of GW interferometers to the ULDM fluctuations.

\begin{figure}[t]
\centering
\includegraphics[width=0.45\textwidth]{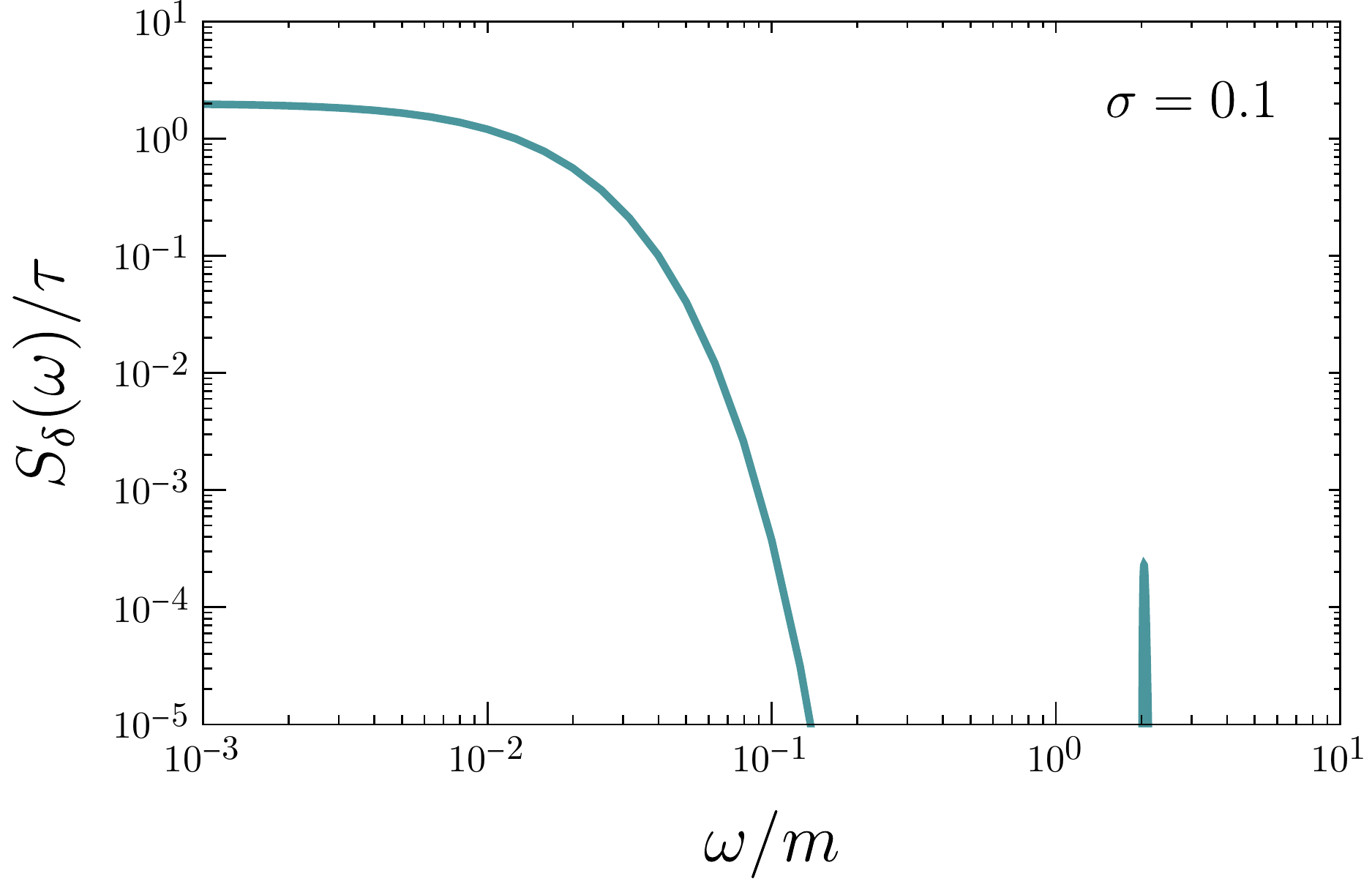}
\quad
\includegraphics[width=0.45\textwidth]{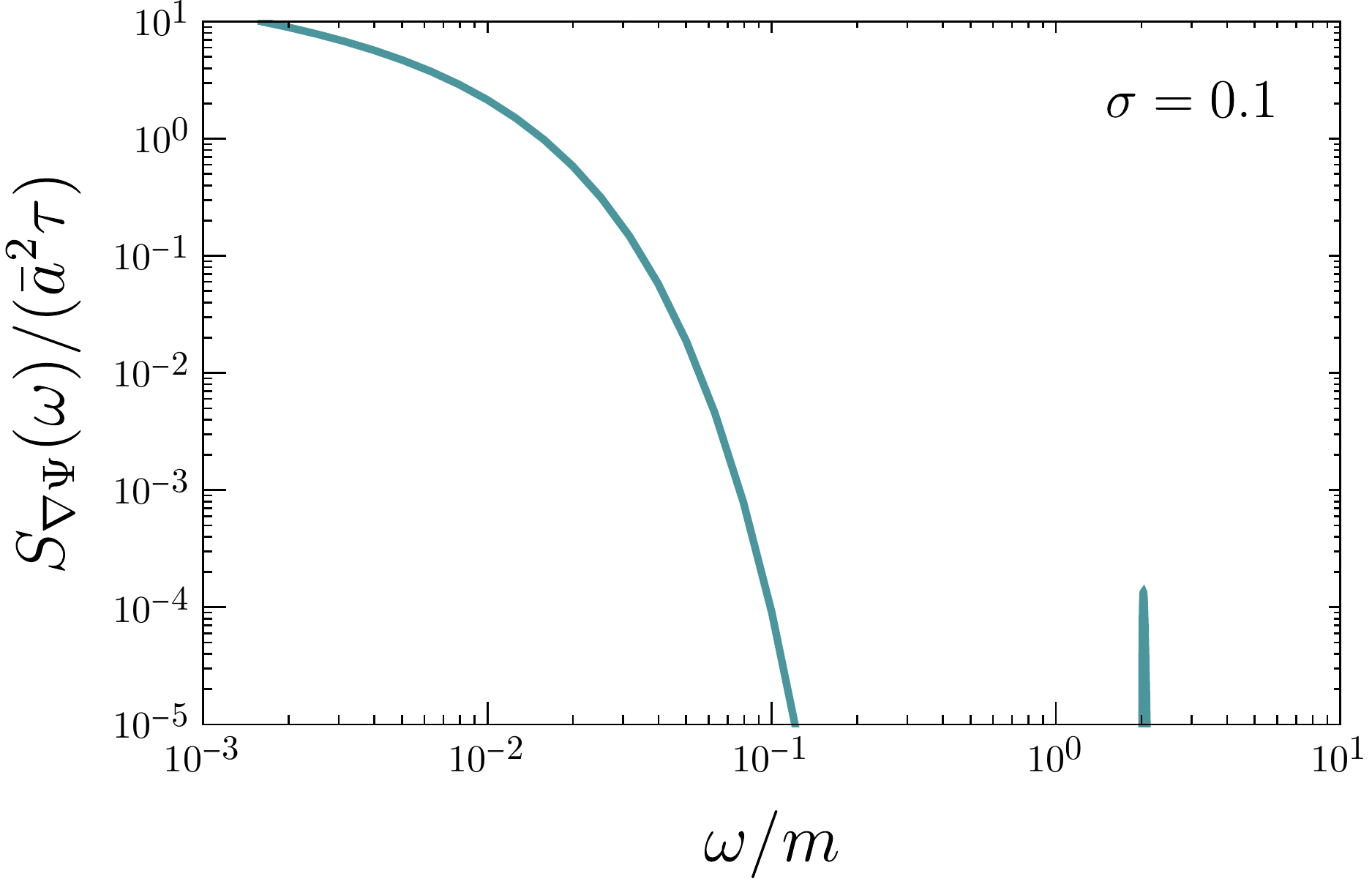}
\caption{(Left) the density contrast spectrum $S_\delta(\omega) = S_\delta(\omega, \vec 0)$ in the isotropic limit $v_0=0$. 
(Right) the spectrum for the gradient of the metric fluctuation, $S_{\nabla \Psi}(\omega) = \delta^{ij} S_{\nabla\Psi}^{ij}(\omega, \vec 0)$. 
Both spectra exhibit two distinctive frequency components: one at $\omega =2m$ and another at $\omega < m \sigma^2$.
For demonstration, we choose unrealistically large velocity dispersion $\sigma = 0.1$. 
The narrow peak at $\omega =2m$ is suppressed by $\sigma^4$ compared to the smooth low-frequency spectrum.
For $S_{\nabla\Psi}$, there is a logarithmic divergence at low frequencies due to the long-range nature of gravitational force. 
}
\label{fig:delta_spectrum}
\end{figure}

To illustrate the statistical properties of density contrast, we explicitly compute $S_\delta(\omega, \vec 0)$. 
We consider the normal distribution for dark matter velocity distribution, 
\bea
f(\vec v) = \frac{\rho_0/m}{(2 \pi \sigma^2)^{3/2}} \exp\left[ - \frac{(\vec v - \vec v_0)^2}{2\sigma^2} \right] , 
\label{normal_dist}
\eea
where $\vec v_0$ is the mean dark matter velocity, and $\sigma$ is the velocity dispersion. 
We work in the rest frame of the GW detectors. 
In general, $\vec v_0\neq0$ in this frame, but for the sake of simplicity, we take the isotropic limit $\vec v_0=0$ for now. 
Normalization of $f(\vec v)$ is fixed by~\eqref{normalization_f}. 
Using \eqref{deltarho_PS} and \eqref{S_P}, we find
\bea
S_{\delta} (\omega, \vec 0)
= 
\tau
\Big[
\sigma^4 A_\delta(\omega) + B_\delta(\omega)
\Big],
\label{spec_delta}
\eea
where $\delta = \delta \rho_\phi / \rho_0$ is density contrast, and $\tau = 1/ m\sigma^2 = 1/\omega_0$ is the coherence time.
In the isotropic limit ($v_0/\sigma \ll 1$), the $A_\delta(\omega)$ and $B_\delta(\omega)$ are given by
\begin{align}
A_\delta(\omega) &= \frac{5\pi}{32} \Big( \frac{ \bar v}{\sigma} \Big)^8 
\exp\Big[ - \frac{\bar v^2}{\sigma^2} \Big] \theta(\bar v^2)
 + ( \omega \to - \omega ) , 
\\
B_\delta(\omega) &= 2 |\bar\omega| K_1(|\bar\omega|) , 
\end{align}
where $K_n(x)$ is the modified Bessel function.
Here we introduce, for notational simplicity,
\bea
\bar v^2(\omega) = (\omega/m -2),
\quad \textrm{and} \quad
\bar\omega= \omega / (m\sigma^2) = \omega / \omega_0 .
\label{def_misc}
\eea
The spectrum $S_\delta(\omega, \vec 0)$ for $\vec v_0 =0$ and $\sigma =0.1$ is shown in the left panel of Figure~\ref{fig:delta_spectrum}. 

This computation reveals an interesting feature in the ULDM density fluctuation.
The spectrum contains two distinctive frequency components: a narrow peak at $\omega=2m$ represented by $A_\delta(\omega)$ and a broad low-frequency spectrum at $\omega < m \sigma^2$ represented by $B_\delta(\omega)$.
Both of them have a width $\Delta \omega \sim m \sigma^2$. 
They can be understood by investigating the energy density of a single mode, $\phi(t,\vec x) = \phi_0 \cos(\omega t - \vec k \cdot \vec x)$,
$$
\rho_\phi = \frac{1}{2} 
\Big[
\dot{\phi}^2 
+ (\nabla \phi)^2
+ m^2 \phi^2
\Big]
\approx \rho_0 
\Big[
1 - v^2 \cos(2 m t + \varphi )
\Big].
$$
It consists of dc mode, which is what constitutes the low-frequency part ($\omega < m\sigma^2$) for a realistic multi-mode density contrast, and a coherently oscillating part which constitutes a spectrum at $\omega =2m$ with its amplitude suppressed by $v^2$.
This $v^2$ suppression is responsible for a relative suppression of $\sigma^4$ in front of $A_\delta(\omega)$ in~\eqref{spec_delta}. 
These two distinctive frequency components in the density contrast remain in the ULDM-induced signal/noise power spectrum for the gravitational detectors, offering multiple ways to search for ultralight dark matter at a different mass range.

\subsection{Metric fluctuation}
ULDM fluctuations source metric perturbations.
The metric is given by
\bea
ds^2 = ( 1 + 2 \Phi) dt^2 - (1 - 2\Psi) dx^2. 
\label{metric_perturbation}
\eea
These metric fluctuations perturb the position of test masses in interferometers and induce a time delay of light. 
Suppose a test mass at the position $\vec x$.
The test mass acceleration and the metric fluctuations satisfy
\begin{align}
\vec a &= - \nabla \Phi , 
\label{a}
\\
\nabla^2 \Psi &= 4\pi G \delta\rho_\phi,
\label{field_eq1}
\\
6\ddot{\Psi} + 2 \nabla^2 ( \Phi - \Psi) &= 24\pi G \delta P_\phi.
\label{field_eq2}
\end{align}
Here $\delta P_\phi$ is the pressure perturbation of the scalar field $\phi$.
The pressure of the field is $P_\phi = \dot{\phi}^2/2 - (\nabla\phi)^2/6 - m^2\phi^2/2$. 
Statistical properties of metric fluctuation as well as test mass position can be derived from those of ULDM by solving the above equations. 

The relation between test mass acceleration, metric perturbations, and ULDM fluctuations becomes transparent in Fourier space:
\begin{align}
\tilde a^i(k) &= - i  k^i \tilde \Phi(k) , 
\label{a_Fourier}
\\
\tilde \Psi(k) &=  - \frac{4\pi G}{k^2} \widetilde{\delta\rho}_\phi(k) , 
\label{Psi_Fourier}
\\
\tilde \Phi(k) &= - \frac{4\pi G}{k^2} 
\left[
\Big( 1 - \frac{3\omega^2}{k^2} \Big) \widetilde{\delta\rho}_\phi(k)
+ 3 \widetilde{\delta P}_\phi(k) \right] .
\label{Phi_Fourier}
\end{align}
From above, one can derive the relation between power spectra, e.g. $P_\Psi(k) = ( 4\pi G/k^2 )^2  P_{\delta \rho} (k)$ and $P_a^{ij}(k) = k^i k^j P_\Phi(k)$, where the test mass acceleration power spectrum is defined as $\langle \tilde a^i(k) \tilde a^{j*}(k') \rangle = (2\pi)^4 \delta^{(4)}(k-k') P^{ij}_a(k)$.

Among others, the power spectrum of $\nabla \Psi$ turns out to be particularly useful for the later discussion. 
In the frequency space, it is given by
\bea
S_{\nabla \Psi}^{ij}(\omega , \vec L)
= \int \frac{d^3k}{(2\pi)^3} e^{i \vec k \cdot \vec L } k^i k^j P_\Psi(k)
= \bar a^2 \tau \big[ \sigma^4 A^{ij} + B^{ij} \big] 
\label{S_dP}
\eea
where $A^{ij}$ and $B^{ij}$ are 
\begin{align}
A^{ij} (\omega, \vec L)
&=
16 \Big( \frac{\bar v}{\sigma} \Big)^6
\exp\Big[ - \frac{v_0^2 + \bar v^2}{\sigma^2} \Big]
\theta(\bar v^2)
\left[
\delta^{ij} \frac{I_3(X)}{X^3}
+  X^i X^j \frac{I_4(X)}{X^4} 
\right]
+ ( \omega \to - \omega ) , 
\\
B^{ij}(\omega, \vec L) &= 
\frac{2}{\sqrt{\pi}} \int_{-\infty}^\infty ds \, \frac{e^{-i s \bar \omega}}{\sqrt{1+s^2}} 
\left[ 
\delta^{ij}  \frac{{\rm erf}(Y) - G(Y) }{Y}
+ \frac{Y^i Y^j}{Y^3} \big( 3 G(Y) -{\rm erf}(Y) \big)
\right], 
\end{align}
with $G(X) = [1/(2X^2)]({\rm erf}(X) - 2 X e^{-X^2} / \sqrt{\pi} )$ and
\begin{align}
\vec X = 2 \frac{\bar v }{\sigma } 
\Big( \frac{\vec v_0}{\sigma} + i \vec L_\lambda \Big),
\qquad
\vec Y = \frac{1}{\sqrt{1+s^2}} 
\Big(\frac{s \vec v_0}{\sigma} + \vec L_\lambda \Big),
\qquad
\vec L_\lambda = m \sigma \vec L . 
\end{align}
The $\bar v$ and $\bar \omega$ are defined in~\eqref{def_misc}. 
Here $I_n$ is the modified Bessel function of the first kind.
Recall that  $\tau  = 1 / m\sigma^2$, $m_{\rm eff} =  \pi^{3/2} \rho_0 / (m\sigma)^3$, and $\lambda = (m\sigma)^{-1}$ and the typical acceleration $\bar a =G m_{\rm eff} / \lambda^2$ .
We will see shortly that the response of the GW interferometer depends mostly on $\nabla \Psi$, and the above power spectrum can be used as a basic building block to study the response of interferometers to the ultralight dark matter fluctuations.

The structure of the spectrum is similar to the density contrast spectrum \eqref{spec_delta} since $\Psi$ inherits its properties from ULDM fluctuations. 
Especially, $A^{ij}(\omega, \vec L)$ is centered at $\omega =2m$, while $B^{ij}(\omega,\vec L)$ is spread over $\omega \lesssim m\sigma^2$. 
In the isotropic ($v_0 = 0$) and the long-wavelength limit ($m\sigma L \ll 1$), the $A^{ij}$ and $B^{ij}$ are simplified to
\begin{align}
A_0^{ij} (\omega)
&\approx
\frac{1}{3}
\Big( \frac{\bar v}{\sigma} \Big)^6
\exp\Big[ - \frac{\bar v^2}{\sigma^2} \Big]
\theta(\bar v^2)
\Big[
\delta^{ij}
- \frac{(m \bar v L)^2}{4} ( \delta^{ij} + 2 \hat L^i \hat L^j )
\Big]
+ ( \omega \to - \omega ) , 
\\
B_0^{ij}(\omega) &\approx
\frac{16}{3 \pi} 
\Big[
\delta^{ij} K_0( | \bar\omega| ) 
- \frac{( m \sigma L)^2}{5} (\delta^{ij} +2 \hat L^i \hat L^j ) |\bar\omega| K_1( |\bar\omega| )
\Big].
\end{align}
The spectrum $S_{\nabla \Psi}(\omega) = \delta^{ij} S_{\nabla\Psi}^{ij}(\omega, \vec 0)$ is shown in the right panel of Figure~\ref{fig:delta_spectrum}. 
Naively, this can be considered as a test mass acceleration power spectrum. 
While the overall structure of the spectrum is similar, the acceleration spectrum logarithmically diverges at small $\omega$.
More specifically, $B_0^{ij} \propto \delta^{ij} \ln(1/\bar\omega)$ for $\bar\omega \ll1$, and this logarithmic divergence is related to the long-range nature of the gravitational interaction. 
The derivation of $B^{ij}(\omega, \vec L)$ is given in Appendix~\ref{App:PS}.

\section{Gravitational wave interferometers}\label{sec:GW_int}

\begin{figure}
\centering
\includegraphics[width=0.6\textwidth]{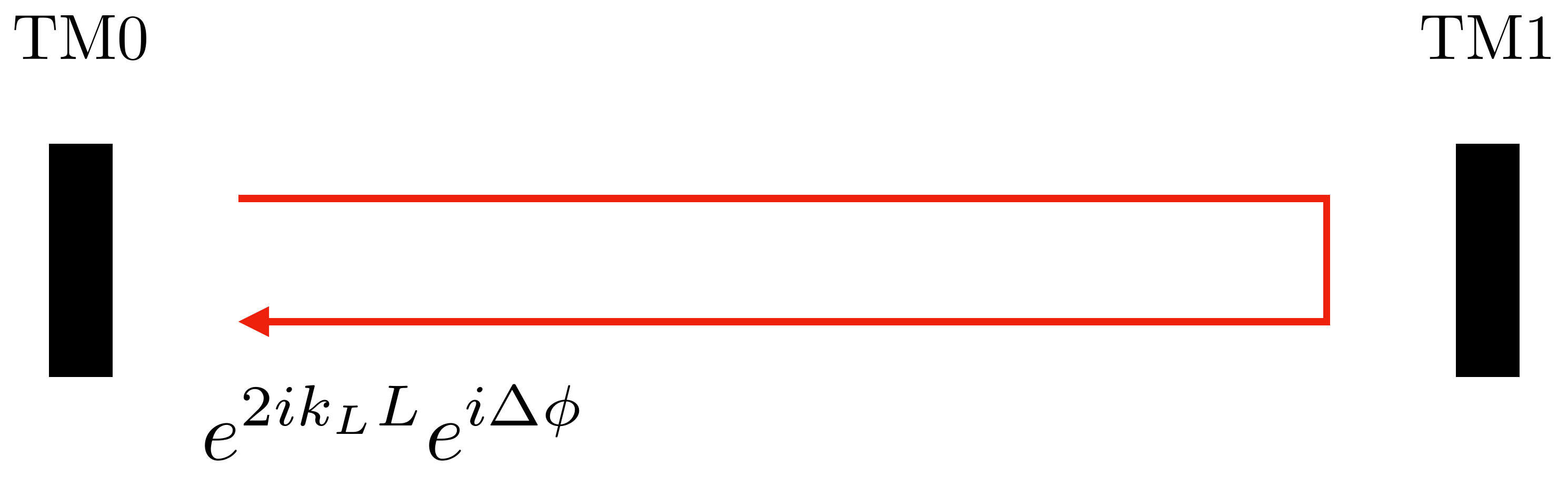}
\caption{The laser begins at $t_0$ at TM0. It propagates to TM1 and returns to TM0 at time $t$. 
During one round-trip, the laser acquires a phase $e^{2 i k_L L}$ without ultralight dark matter. 
With ultralight dark matter, additional phase $e^{i\Delta\phi}= e^{2i\omega_L(\delta t + \delta L)}$ is induced because of (i) time-delay $\delta t$ of the laser light and (ii) the fluctuation of the distance between two test masses $\delta L$. }
\label{fig:TM0TM1}
\end{figure}

To investigate how each interferometer responds to the ULDM fluctuation, we first consider a simple setup shown in Figure~\ref{fig:TM0TM1} where a phase-coherent laser begins from TM0, reaches TM1, and returns to TM0 again at $t$. 
The phase of the laser at time $t$ is given by
$$
e^{- i \omega_L t_0 }
= 
e^{- i \omega_L t }  e^{2 i k_L L} e^{i\Delta \phi},
$$
since the phase along the null trajectory does not change. 
Here $(\omega_L, \vec k_L)$ is the laser frequency and wave vector, and $t_0$ is the time at which the laser begins from TM0. 
Since $t_0 = t - 2L$, the laser acquires additional phase $e^{2 i k_L L}$ during this one round-trip.

Ultralight dark matter introduces an additional phase in two different ways: it introduces the time delay of laser light and it perturbs the test mass position. 
The additional phase $e^{i\Delta \phi}$ is introduced as a result, and it is given by
\bea
\Delta \phi (t) = 2 \omega_L ( \delta t + \delta L)
\label{phase_shift}
\eea
where the test mass perturbation $\delta L$ and the time delay $\delta t$ are given as
\begin{align}
\delta L & = 
+ \frac{1}{2} \hat x \cdot 
\Big[ 
\big( \delta \vec x_1( t- L) - \delta \vec x_0 (t-2L) \big)
+ \big( \delta \vec x_1( t- L) - \delta \vec x_0 (t) \big) 
\Big], 
\label{l_pert}
\\
\delta t & = - \frac{1}{2}  \left[ \int^{t}_{t-L} dt' + \int^{t-L}_{t-2L} dt' \right] 
[ \Psi(t',\vec x(t')) + \Phi(t',\vec x(t'))  ] ,
\label{t_delay}
\end{align}
where $\delta \vec x_0$ and $\delta \vec x_{1}$ are the perturbation of the position of TM0 and TM1 around their nominal positions, $\vec x_0$ and $\vec x_{1}$. 
Note that the argument $\vec x(t')$ in the metric perturbation must be computed along the null trajectory. 
The above expression is obtained by solving null geodesic with the metric perturbation; specifically, one finds $t_0 = t - 2 (L + \delta L) - 2 \delta t$.

The additional phase $\Delta\phi$ can be expanded in terms of metric perturbations in the Fourier space. 
In particular, it can be approximated as
\bea
\widetilde{\Delta \phi}(\omega)
\approx \pm \frac{\omega_L}{\omega^2} e^{i \omega L}
\int \frac{d^3k}{(2\pi)^3} e^{i\vec k \cdot \vec x_0} 
\big[ i \vec k \cdot \vec D(\hat n, L) \big]
\tilde\Psi(k),
\label{D_phi}
\eea
where $\hat n = (\vec x_1 - \vec x_0)/L$ and the vector $\vec D(\hat n, L) = 2 \hat n [ e^{i \vec k \cdot \hat n L} - \cos(\omega L) ] $ encodes the response of this simple system to the ULDM fluctuations. 
The detailed derivation of the expression is given in Appendix~\ref{app:phase}. 

\subsection{Interferometry response}
\subsubsection{Michelson interferometer}
Consider now the Michelson interferometer. 
The Michelson interferometer consists of two arms.
The laser combines at the asymmetric output, and the resulting power is $P = P_0 \sin^2( k_L \Delta L + \Delta \phi_{\rm Mich})$ where $\Delta \phi_{\rm Mich} = \frac{1}{2} ( \Delta \phi_x - \Delta \phi_y)$ with the additional phase $\Delta \phi_{x,y}$ acquired by the laser that has propagated along $x-$ and $y-$arm, respectively. 
The expression for $\Delta \phi_{x,y}$ is given in \eqref{D_phi}. 
In the frequency space, the phase is
\bea
\widetilde{\Delta \phi}_{\rm Mich}(\omega) 
= \pm \frac{\omega_L}{\omega^2} e^{i\omega L}
\int \frac{d^3k}{(2\pi)^3} 
e^{i \vec k \cdot \vec x_0} 
\big[ i \vec k \cdot \vec D_{\rm Mich}(\hat x ,\hat y, L)
\big] \tilde \Psi(k)
\eea
where $\vec D_{\rm Mich}(\hat x, \hat y; L) = \frac{1}{2} [ \vec D(\hat x, L) - \vec D(\hat y, L) ]$ is the response vector of the Michelson interferometer, and $\vec x_0$ is the beam splitter position. 
For simplicity, we assume an equal arm-length. 

The power spectrum for differential arm-length $\widetilde{\Delta L} / L = \widetilde{\Delta\phi}_{\rm Mich} / (\omega_L L)$ can be directly obtained from the above expression, and it is given by
\bea
S^{\rm ULDM}_{\Delta L / L}(\omega) = 
\frac{1}{\omega^4 L^2}
\int \frac{d^3k}{(2\pi)^3} | \vec k \cdot \vec D_{\rm Mich}|^2 P_\Psi(k)
= \frac{\bar a^2}{\omega^4 L^2} \tau 
\Big[ \sigma^4 A_{\rm Mich} + B_{\rm Mich} \Big] . 
\eea
The detector response is encoded in the function $A_{\rm Mich}$ and $B_{\rm Mich}$.
Note again the similar structure of the power spectrum compared to the density contrast and metric fluctuation $\nabla\Psi$. 
The $A_{\rm Mich}$ and $B_{\rm Mich}$ can be decomposed in terms of $A^{ij}$ and $B^{ij}$ in \eqref{S_dP} as follows:
\begin{align}
A_{\rm Mich}(\omega; \hat x,\hat y , L) = &
2 \Big[
\frac{ 1+\cos^2(\omega L)}{2} \big( A_R^{11}(\vec 0) + A_R^{22}(\vec 0)  \big)
- \cos(\omega L) \big( A_R^{11}(\vec x_1) + A_R^{22}(\vec x_2) \big)
\nonumber\\
& - \big( 
A^{12}_R(\vec x_{12}) + \cos^2(\omega L) A^{12}_R(\vec 0) 
- \cos(\omega L) ( A^{12}_R(\vec x_1) + A^{12}_R(\vec x_2) ) 
\big)
\Big], 
\\
B_{\rm Mich}(\omega; \hat x,\hat y , L) = &
2 \Big[
\frac{ 1+\cos^2(\omega L)}{2} \big( B_R^{11}(\vec 0) + B_R^{22}(\vec 0)  \big)
- \cos(\omega L) \big( B_R^{11}(\vec x_1) + B_R^{22}(\vec x_2) \big)
\nonumber\\
& - \big( 
B^{12}_R(\vec x_{12}) + \cos^2(\omega L) B^{12}_R(\vec 0) 
- \cos(\omega L) ( B^{12}_R(\vec x_1) + B^{12}_R(\vec x_2) ) 
\big)
\Big], 
\end{align}
where $A_R^{ab} = \hat x_a^i \hat x_b^j A_R^{ij}$, $B_R^{ab} = \hat x_a^i \hat x_b^j B_R^{ij}$, 
$$
A_R^{ij} (\vec x) = \frac{1}{2} \Big[ A^{ij}(\vec x)  + A^{ij}(-\vec x) \Big],
\quad
\textrm{and}
\quad
B_R^{ij} (\vec x) = \frac{1}{2} \Big[ B^{ij}(\vec x)  + B^{ij}(-\vec x) \Big]. 
$$
Here $(\hat x_1, \hat x_2) = (\hat x,\hat y)$ denote the unit vector to each arm. 
The dependence of $A_R^{ab}$ on $\omega$ is suppressed in the above expression for notational simplicity. 
This power spectrum can be compared with the noise power spectrum in the current and future ground-based interferometers, such as LIGO and Einstein Telescope. 
While we only discuss the Michelson interferometer here, the response of Fabry-P{\'e}rot Michelson interferometer is the same up to the overall multiplication factor.

The response functions can be simplified in the isotropic ($v_0/\sigma \ll 1$) and long-wavelength limit ($m \sigma L \ll 1$).
In such limits, they become
\begin{align}
A_{\rm Mich} &\approx
\frac{8}{3} \Big( \frac{\bar v}{\sigma} \Big)^6
\theta(\bar v^2)
e^{- (\bar v/\sigma)^2}
\Big[ 
\sin^4 \Big( \frac{\omega L}{2} \Big) 
+ \frac{(m \bar v L)^2}{4} 
\Big( \! \cos\omega L - \sin^2\Big(\frac{\omega L}{2}\Big) \Big)
\Big]
+ (\omega \to -\omega) ,
\\
B_{\rm Mich} &\approx
\frac{128}{15 \pi}
|\bar\omega| K_1( |\bar\omega| )
(m \sigma L)^2. 
\end{align}
In this limit, one sees $B_{\rm Mich} \propto (m \sigma L)^2 = (L / \lambda)^2$, which signals the tidal limit of differential acceleration as estimated in the introduction. 
On the other hand, for $A_{\rm Mich}$, the prefactor strongly supports $\omega =2m$. 
The first term $\sin^4(\omega L/2)$ dominates the second term as long as $m \sigma L \gtrsim \sigma^2$. 

\begin{figure}
\centering
\includegraphics[width=0.4\textwidth]{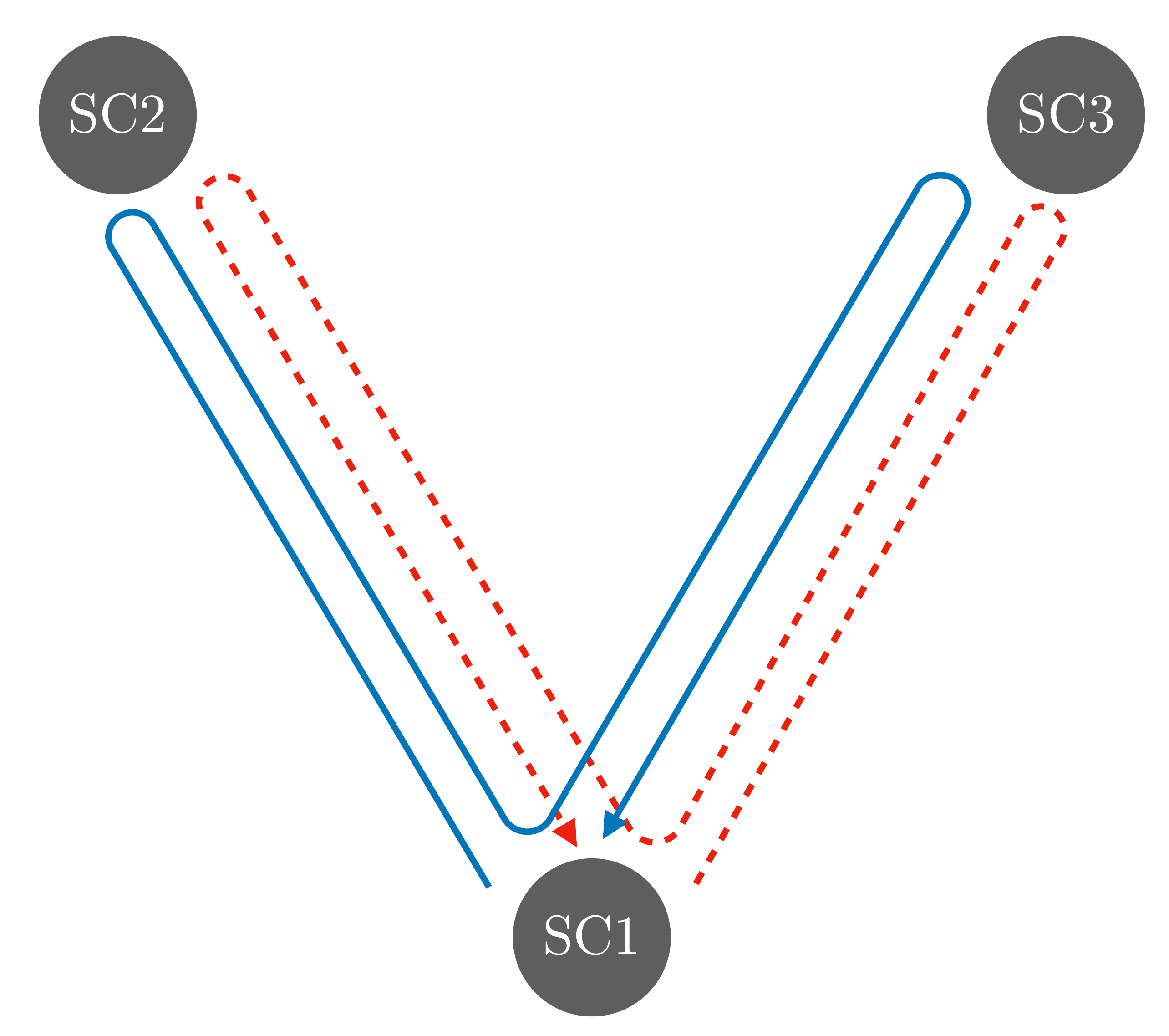}
\caption{A cartoon for the time-delay interferometer. 
Each spacecraft is loaded with two proof masses. 
The Michelson-like X variable measures the phase difference of the laser along two paths shown as blue and red dashed lines. }
\label{fig:TDI_cartoon}
\end{figure}

\subsubsection{Time-delay interferometer}
Let us now consider the time-delay interferometer.
We take LISA as a benchmark for the discussion. 
The constellation consists of three spacecraft (SC$_1$, SC$_2$, SC$_3$) with three arms, and in each arm, two Doppler shift measurements are carried out; in this simplified picture, there are six one-way Doppler shift measurements.
For a detailed discussion of the time-delay interferometer, we refer readers to Refs.~\cite{Otto:2015erp, Tinto:2020fcc}. 

A single one-way Doppler shift measurement can be used for the detection of gravitational waves, but the sensitivity of such measurement will be limited by laser frequency noise.
Such laser frequency noise can be eliminated in a certain linear combination of the Doppler shift measurements; such linear combinations are called time-delay interferometer (TDI) variables~\cite{1999PhRvD..59j2003T, 1999ApJ...527..814A}.

One such TDI variable is called the Michelson-like TDI variable $X$. 
It is defined as
\begin{align}
X(t) 
& = [ \Delta\phi_3(t) + \Delta\phi_2(t-2L) ] - [\Delta\phi_2(t) + \Delta \phi_3(t-2L)] 
\label{TDI_X}
\end{align}
where $\Delta\phi_{2}(t)$ [$\Delta\phi_3(t)$] is the phase obtained by the laser along its round trip from SC1 to SC2 (SC3) at time $t$. 
This TDI-X variable measures the laser phase difference between two paths shown in Figure~\ref{fig:TDI_cartoon}. 
Assuming an equal arm-length for simplicity, the $X$ in the Fourier space is\footnote{In each spacecraft, two proof masses are loaded. We denote the position of proof masses loaded in SC$_i$ as $\vec x_i$. 
We do not distinguish the difference in the position of two proof masses loaded in the same spacecraft.
This is equivalent to assuming that the ULDM fluctuation perturbs two proof masses in the same spacecraft in the same way. This can be justified by noting that the wavelength of dark matter is always larger than the separation of proof masses in the same spacecraft.
}
\bea
\tilde X(\omega) = \pm \frac{\omega_L}{\omega^2} e^{i\omega L}
\int \frac{d^3k}{(2\pi)^3}
e^{i \vec k \cdot \vec x_1} 
\big[ i \vec k \cdot \vec D_{X}(\hat n_3,\hat n_2; L) \big]
\tilde \Psi. 
\eea
where
\bea
\vec D_{X} ( \hat n_{3}, \hat n_{2}; L)
= (1 - e^{2 i \omega L} ) [ \vec D(\hat n_{3}, L) - \vec D(\hat n_{2}, L) ]
= 2 (1 - e^{2 i \omega L} ) \vec D_{\rm Mich} ( \hat n_{3}, \hat n_{2}; L). 
\eea
Here $\hat n_{2}$ ($\hat n_3$) is the unit vector from SC$_1$ and SC$_2$ (SC$_3$). 
The response vector $\vec D_{X}$ is the same as that of the Michelson interferometer except for an overall factor of $2(1- e^{2i\omega L})$. 
In the computations of the power spectrum, this additional factor leads to a common multiplication factor $16\sin^2(\omega L)$, which will be ignored below.

The power spectrum due to ultralight dark matter fluctuation is
\bea
S^{\rm ULDM}_{\Delta L / L}(\omega) 
= \frac{\bar a^2}{\omega^4 L^2} \tau 
\Big[
\sigma^4 A_{\rm Mich}(\omega; \hat n_3,\hat n_2, L) 
+ B_{\rm Mich}(\omega; \hat n_3,\hat n_2, L) 
 \Big]
\eea
which is the same as the simple Michelson interferometer except that the two arms are not orthogonal to each other, $\hat n_2 \cdot \hat n_3 = 1/2 \neq0$. 
The noise power spectrum is~\cite{Babak:2021mhe}
\bea
S_{\Delta L / L}( \omega ) = 
\frac{1}{L^2}
\left[ S_N( \omega ) + 2 [ 1 + \cos^2(\omega L) ] \frac{S_{\rm acc} (\omega)}{\omega^4} \right] .
\label{PS_X_TDI}
\eea
where $S_N(f)$ is the  optical metrology noise power spectrum, and $S_{\rm acc}(\omega)$ is the acceleration noise power spectrum.
In this expression, the optical metrology and acceleration noise in each spacecraft are assumed to be identical, $S_{N_{ij}} = S_N$ and $S_{{\rm acc},ij} = S_{\rm acc}$. 
The noise spectrum and the ULDM spectrum will be compared later.

Similar to the Michelson interferometer, the response function can be obtained analytically in the isotropic and long wavelength limit.
Assuming an equilateral configuration, we find
\begin{align}
A_X &\approx 
\frac{4}{3} \Big( \frac{\bar v}{\sigma} \Big)^6
e^{-(\bar v / \sigma)^2} 
\theta(\bar v^2)
\Big[ 
\sin^4 \Big(\frac{\omega L}{2} \Big) + \frac{3}{8} (m \bar v L)^2 \cos(\omega L)
\Big]
+ (\omega \to -\omega)
\\
B_X & \approx 
\frac{32}{5\pi} 
|\bar\omega| K_1(|\bar\omega|)
(m \sigma L)^2
\end{align}
The behavior of the response function in the time-delay interferometer is the same as in the Michelson interferometer. 

\subsection{ULDM signal}
Two distinct frequency components in the ULDM spectrum allow us to perform two different sets of analyses for the ultralight dark matter search. 
In what follows, we discuss how each frequency component of ultralight dark matter can be searched with matched filtering and cross-correlation of multiple detector outputs. 

\subsubsection{Deterministic signal}
The $\omega =2m$ part of the spectrum represents the coherently oscillating ULDM signals.
This is a deterministic signal in the sense that the signal form is known. 
For such deterministic signals, we can use the matched filter.
Consider the detector output $d(t) = s(t) + n(t)$. 
By convoluting the detector output with the filter, $\int dt \, d(t) K(t)$, and choosing the filter  $K(t)$ such that it maximally overlaps with the signal $s(t)$, one can add the signal coherently.
The signal-to-noise ratio is
\bea
\frac{S}{N} =
\left[ 
\min(T, \sqrt{T\tau} )
\int_{-\infty}^{\infty} df \, \frac{S_s(f)}{S_n(f)} 
\right]^{1/2}
\eea
where $T$ is the total integration time scale and $\tau = 1 / m\sigma^2$ is the coherence time scale of the deterministic signal. 
Here $S_s(f)$ is the power spectrum in differential arm length or strain unit from the ULDM fluctuations. 
The factor $\min(T, \sqrt{T \tau})$ denotes the gain in the signal-to-noise ratio in the coherent search, $S/N \sim \sqrt{T}$ when $T < \tau$ and the gain in the incoherent search, $S/N \sim T^{1/4}$ when $T > \tau$. 
This will be used to place an upper limit on dark matter density in the solar system.  
Note that the power spectrum in this expression is defined as a two-sided spectrum. 

To proceed further, we note that the power spectrum of the coherently oscillating mode is narrowly peaked at $\omega =2m$ with a width $\Delta \omega = m \sigma^2$.
For the Michelson interferometry, the integral can be approximated as
\bea
\frac{S}{N} \approx
\frac{\bar a \sigma^2}{2\pi^2 f_m^2 L}
\bigg[ 
\frac{\min(T, \sqrt{T\tau} )}{S^{\rm one}_{\Delta L/L}(f_m)} 
\bigg]^{\frac12}
\bigg[
\tau \int_{0}^{\infty} df \, A_{\rm Mich}(f) 
\bigg]^{\frac12}
\eea
where $f_m = m/\pi$ and $S^{\rm one}_{\Delta L /L}(f) = 2 S_{\Delta L / L}(f)$ is the one-sided noise power spectrum. 
We have evaluated all the other factors in the integral at $f= f_m$ except for the exponential factor strongly peaked at $\omega=2m$, assuming that they are smoothly distributed over $\Delta \omega \simeq m \sigma^2$. 
In the isotropic ($v_0/\sigma \ll 1$) and long-wavelength limit $(m \sigma L \ll 1)$, one finds $\tau \int_0^\infty df \, {\cal A}_{\rm Mich} \simeq (8/\pi) \sin^4(f_m/2f_*)$ with $f_* = (2\pi L)^{-1}$. 
The above expression simplified to
\bea
\frac{S}{N} =
\frac{2^{5/2}}{\sqrt{\pi}}
\sin^2(f_m/2f_*) 
\left[
\frac{\bar a \sigma^2}{(2\pi f_m)^2 L}
\right]
\bigg[ 
\frac{\min(T, \sqrt{T\tau} )}{S^{\rm one}_{\Delta L /L}(f_m)} 
\bigg]^{\frac12}.
\label{SNR_A_Mich}
\eea
The quantity in the first squared parenthesis is the typical displacement $\Delta L / L$ due to ULDM fluctuation during $\Delta t \sim 1/f_m$.
Note that $\bar a = G m_{\rm eff}/\lambda^2$ and $m_{\rm eff} = \pi^{3/2} \rho \lambda^3$. The above equation can be solved for $\rho/\rho_0$ with the local dark matter density $\rho_0 = 0.4\,{\rm GeV/cm^3}$. 
For the interferometers with equilateral geometry, $\tau \int_0^\infty A_{\rm Mich} \approx (4/\pi) \sin^4(f_m/2f_*)$. 
The additional factor of $1/2$ is because the two arms are not orthogonal to each other.

\subsubsection{Stochastic signal}
The low-frequency part of the ULDM spectrum behaves similarly to the stochastic gravitational wave background. 
To search for this part of the spectrum, we ideally need more than two detectors, which will allow us to cross-correlate the outputs in different detectors.
If the separation of detectors is smaller than the wavelength of dark matter, the output at each detector is expected to be correlated, while the noise is expected to be uncorrelated. 
In this way, we can single out the ultralight dark matter signal in the output data stream of multiple detectors. 

Consider two detector output $d_{1,2}(t) = s_{1,2}(t) + n_{1,2}(t)$ with the signal $s_{1,2}$ and the noise $n_{1,2}$. 
Then define the cross-correlated output $Y$ as
\bea
Y
= \int_{-T/2}^{T/2} dt \int_{-T/2}^{T/2} dt' \,  d_1(t) d_2(t') Q(t-t')
\simeq \int_{-\infty}^{\infty} df\,  \tilde d_1(f) \tilde d_2^*(f) \tilde Q^*(f)
\eea
where $Q(t)$ is a real filter function. 
Suppose that the cross-correlation of signals at two detectors is
$\langle \tilde s_1(f)  \tilde s_2^*(f')  \rangle = \delta(f-f') S_{12}(f)$. 
Using the optimal filter $\tilde Q(f) = S_{12}(f)/ S_n^2(f)$ and assuming uncorrelated detector noise, one finds the signal-to-noise ratio as~\cite{Maggiore:2007ulw}
\bea
\frac{S}{N} = 
\left[ T \int_{-\infty}^\infty df\, \frac{| S_{12}(f)|^2}{S_n^2(f)} \right]^{1/2}, 
\eea
where $T$ is the total integration time scale. 
In this way, the low-frequency part of the ultralight dark matter signal can be searched. 

For cross-correlation, we are mostly interested in space-borne interferometers for the following reasons. 
For the cross-correlation to be relevant, two conditions need to be met: (i) the ultralight dark matter-induced signal must lie within the detector bandwidth, and (ii) the ultralight dark matter signals must be correlated in two detectors. 
Each interferometer has a range of frequencies to which the detector is sensitive. 
For the LIGO, the minimum frequency that LIGO is sensitive to is about $f_{\rm min} \sim 10\,{\rm Hz}$.
On the other hand, two LIGO detectors, one at Hanford and the other at Livingston, are separated by $L_{\rm det} \simeq 3000\,{\rm km}$. 
To guarantee that the ultralight dark matter signal is correlated over these two detectors, we require $L_{\rm det} < (1/\Delta p) = (\sigma/\omega)$.
This results in $\omega < 5\times 10^{-2}\,{\rm Hz}$, which is more than two orders of magnitude smaller than the minimum detectable frequency in LIGO detectors.
In other words, for the frequency range where the ultralight dark matter signals would be correlated over two LIGO detectors, LIGO detectors are not sensitive to ULDM signals due to a large seismic noise. 

The situation in the space-borne interferometer is slightly different. 
For the sake of discussion, let us consider two co-planar LISA-like constellations. 
The minimum frequency of the LISA detector is $f_{\rm min} \sim 10^{-5}\,{\rm Hz}$.
On the other hand, the detector separation is $L_{\rm det} \sim L =2.5\times 10^6\,{\rm km}$ or less. 
This translates into the condition $\omega < \sigma / L_{\rm det} \simeq 10^{-4}\,{\rm Hz}$ for ULDM signals to be correlated over two detectors. 
There is a range of frequencies where the signals are correlated over detectors, while the detector maintains its sensitivity to the ultralight dark matter signals.
The same arguments hold for other types of proposed space-borne interferometers, such as $\mu$Ares~\cite{Sesana:2019vho}, a proposal using asteroids as GW detectors~\cite{Fedderke:2021kuy}, and Big Bang Observatory (BBO)~\cite{BBO}.

For this reason, we only consider space-borne interferometers. 
The cross-correlation of TDI-X variables is
\bea
S_{12}(f)
= \frac{\bar a^2 \tau}{(2\pi f)^4 L^2} {\cal B}_{\rm cross} .
\eea
The overlap function is
\begin{align}
{\cal B}_{\rm cross} =& 
\cos^2 (2 \pi f L) {\cal B}^{11'}(\vec x_{11'})
+ {\cal B}^{33'}( \vec x_{33'} )
+ {\cal B}^{22'}( \vec x_{22'} )
- {\cal B}^{23'}(\vec x_{23'} ) 
- {\cal B}^{32'}(\vec x_{32'} ) 
\nonumber\\
&
- \cos (2 \pi f L)
\Big[
{\cal B}^{31'}(\vec x_{31'}) + {\cal B}^{13'}(\vec x_{13'})
- {\cal B}^{21'}(\vec x_{21'}) - {\cal B}^{12'}(\vec x_{12'})
\Big], 
\label{BX2}
\end{align}
where $\vec x_i$ is the position of $i$-th spacecraft in the first constellation, 
$\hat n_{2,3} = (\vec x_{2,3} - \vec x_1)/L$, $\hat n_1 = (\vec x_3 - \vec x_2)/L$, $\vec x_{ij'} \equiv \vec x_i - \vec x_{j'}$, and ${\cal B}^{ab'} = \hat n^i_a \hat n^j_{b'} {\cal B}^{ij}$.
The primed quantities are for the second constellation. 
We take $L= L'$ for simplicity.
The signal-to-noise ratio is
\bea
\frac{S}{N}
=
\frac{\bar a^2}{(2\pi f_0)^4 L^2} 
\left[ 
\frac{T}{(2\pi)^2} \int_{-\infty}^{\infty} df \,
\Big( \frac{f_0}{f} \Big)^8
\frac{| {\cal B}_{\rm cross}(f)|^2 }{f_0^2 S_{\Delta L /L}^2(f) } 
\right]^{1/2}. 
\label{SNR_cross}
\eea
where $f_0 = \omega_0 / 2\pi = m \sigma^2/2\pi$. 
The prefactor is the squared value of the typical differential length change, $(\Delta L / L)^2$, over $\Delta t \sim 1/ f_0$. 
The integral will be numerically solved.

\subsection{Application}
We apply the discussion in previous sections to the current and proposed gravitational wave interferometers. 

\subsubsection{LIGO}
We take the sensitivity curve of LIGO O3 from Ref.~\cite{aLIGO:2020wna}.
The strain noise spectrum from ultralight dark matter can be found by using $S_{\Delta L / L} (f) = {\rm sinc}^2(2\pi f L) S_h (f)$.
For the LIGO with $f \lesssim 10^4\,{\rm Hz}$, ${\rm sinc}(2\pi fL)\simeq 1$, we have ignored the additional ${\rm sinc}^2(2\pi fL)$ factor. 
The arm length is $L_{\rm LIGO} = 4\,{\rm km}$. 
As we have already estimated in the introduction, the ULDM-induced noise is by many orders of magnitude smaller than the instrumental noise. 
Still, it can provide an upper limit on the dark matter density in the solar system. 
For the search with a matched filter, we use~\eqref{SNR_A_Mich} and rewrite it as
\bea
\frac{\rho}{\rho_0}
\approx \frac{S}{N} 
\sqrt{\frac{\pi}{2}}
\frac{1}{{\rm sinc}^2(f_m/2f_*) }
\left[ \frac{(2\pi f_*)^2 L}{\bar a_0 \sigma^2} \right]
\Big[ \frac{S_{\Delta L / L}^{\rm one}(f_m)}{\min(T,\sqrt{T\tau} )} \Big]^{\frac12}
\label{rho_LIGO}
\eea
where $\bar a_0 = G m_{\rm eff,0}/\lambda^2$, $m_{\rm eff,0} = \pi^{3/2} \rho_0 \lambda^3$ with $\rho_0 = 0.4\,{\rm GeV/cm^3}$, and $f_* = (2\pi L)^{-1} =10^4\,{\rm Hz}$ for LIGO. 
Note that for LIGO, we can also approximate ${\rm sinc}^2(f_m / 2f_*) \simeq 1$. 
For $T=1\,{\rm yr}$, $\sigma =164\,{\rm km/sec}$, and $S/N=1$, the upper limit on the dark matter density $\rho/\rho_0$ is obtained and shown as a solid red line in Figure~\ref{fig:summary}.
Note that the above expression is valid in the long-wavelength limit $m \sigma L \ll1$, which is satisfied not only in LIGO but also in all other ground-based interferometers. 

\subsubsection{Einstein Telescope}
Einstein telescope is a proposed future gravitational wave detector based on a dual-recycled Michelson interferometer.
The arm length is $L_{\rm ET} = 10\,{\rm km}$. 
We assume an equilateral geometry for the Einstein telescope. 
The strain noise power spectrum is taken from~\cite{Hild:2010id}. 
Using the same expression~\eqref{rho_LIGO} with $f_* = (2\pi L_{\rm ET})^{-1} =5\times 10^3\,{\rm Hz}$, we show the upper limit on the density, $\rho/\rho_0$, from Einstein Telescope with the blue solid line in Figure~\ref{fig:summary}. 
We choose the same parameters: $T=1\,{\rm yr}$, $\sigma =164\,{\rm km/sec}$, and $S/N=1$. 

\subsubsection{LISA}
The noise source in LISA is conveniently divided into the collective optical metrology noise $S_N(f)$ and the single test-mass acceleration noise $S_{\rm acc}(f)$. 
Each of them is given by~\cite{LISA:2017pwj}
\begin{align}
S_{N}^{1/2}(f) & = 10^{-11} \frac{\rm m}{\sqrt{\rm Hz}} 
\bigg[1+ \left( \frac{2 \rm mHz}{f} \right)^4 \bigg]^{1/2} ,
\\
S_{\rm acc}^{1/2}(f) &= 3\times 10^{-15} \, \frac{\rm m \, s^{-2}}{\sqrt{\rm Hz}}
\left[ 1 +\left( \frac{0.4\,{\rm mHz}}{f} \right)^2 \right]^{\frac12}
\left[ 1 +\left( \frac{f}{8\,{\rm mHz}} \right)^4 \right]^{\frac12}. 
\end{align}
LISA Pathfinder has reported differential acceleration noise of two test masses loaded in a single spacecraft, which is smaller than the LISA proposal requirement by a factor of few~\cite{Armano:2016bkm, Armano:2018kix}. 
For the matched filter search, we reuse \eqref{rho_LIGO} with an additional factor of $\sqrt{2}$, arising from the fact that LISA has an equilateral geometry and two arms are not orthogonal to each other. 
When recycling \eqref{rho_LIGO}, we replace ${\rm sinc}^2(f_m/2f_*) \to 1/[1+ (f_m/2f_*)^2]$ to show the envelope of the projection. 
The upper limit on the dark matter density $\rho/\rho_0$ with the same parameters is shown as a green solid line in Figure~\ref{fig:summary}. 

For the LISA, we also compare $S_{\Delta L/L}^{\rm ULDM}(f)$ with the mission requirement $S_{\Delta L/L}(f)$. 
For the comparison, we compare two spectra in the GW strain unit. 
To convert to the strain noise power spectrum, we divided the power spectrum for the differential arm-length with the sky and polarization averaged detector response function ${\cal R}(f)$~\cite{Robson:2018ifk},
$$
{\cal R} (f) = \frac{3}{10( 1 + 0.6(f/f_*)^2)}
$$
with $f_* = (2\pi L_{\rm LISA})^{-1} \simeq 19\,{\rm mHz}$. 
In the left panel of Figure~\ref{fig:ULDM_noise}, we show the mission requirement (red solid line) and $S_n^{\rm ULDM}$ with $m = 10^{-12}\,{\rm eV}$ (cyan) and $m = 10^{-13}\eV$ (orange).
The ULDM-induced noise is several orders of magnitude smaller than the mission requirement, and therefore, it can safely be ignored for the purpose of gravitational wave detection. 

The subdominant low-frequency ULDM signals might be probed if there are more than two detectors. 
Although the current LISA mission concept does not invoke more than two constellations~\cite{LISA:2017pwj}, we assume two LISA-like interferometers and investigate the reach of such configuration for the ULDM search as an exercise.
Note that other space-borne interferometers, such as Taiji and TianQin, would allow the cross-correlation searches~\cite{Cai:2023ywp}.
For the cross-correlation, the position of the spacecraft needs to be specified.
In the heliocentric ecliptic coordinate, their position is~\cite{Cornish:2001bb}
\begin{align}
x &= a \cos \alpha + a e [ \sin \alpha \cos \alpha \sin \beta - (1 + \sin^2\alpha) \cos\beta ]
\\
y &= a \sin \alpha + a e [ \sin \alpha \cos \alpha \cos \beta - (1 + \cos^2\alpha) \sin\beta ]
\\
z &= \sqrt{3} a e \cos(\alpha-\beta)
\end{align}
where $a \approx {\rm AU}$ is the semi-major axis, $e$ is eccentricity, 
$\alpha = \omega t + \kappa$, and $\beta = 2 n \pi /3 + \lambda$ with $n=0,1,2$. 
Here $\alpha$ is the orbital phase of the guiding center.
Note that the arm length is given by $L= 2 \sqrt{3} \eps R$; for $R ={\rm AU}$ and $L_{\rm LISA} = 2.5\times 10^6\,{\rm km}$, the eccentricity is $\eps = 4.8\times 10^{-3}$. 
The position of spacecraft in the second constellation is given by the same expression but with a prime ($\kappa'$, $\lambda'$, and so on). 

Rewriting~\eqref{SNR_cross} in terms of $\rho/\rho_0$, we find
\bea
\frac{\rho}{\rho_0} 
= 
\frac{1}{2^{3/4}}
\Big( \frac{S}{N} \Big)^{\frac12}
\left[
\frac{(2\pi f_0)^2 L}{\bar a_0}
\right]
\left[ 
\frac{T}{(2\pi)^2}
\int_0^\infty df \, 
\Big( \frac{f_0}{f}\Big)^8 
\frac{|{\cal B}_{\rm cross}|^2}{f_0^2 |S_{\Delta L / L}^{\rm one}(f) |^2} \right]^{-\frac14}. 
\label{rho_cross}
\eea
Using the halo dark matter parameters, $\sigma =164\,{\rm km/sec}$ and $\rho_0 = 0.4\,{\rm GeV/cm^3}$, and assuming $\Delta \kappa = \kappa' - \kappa =0$ (co-planar configuration), $\Delta \lambda = \lambda' - \lambda = \pi/2$, and $T = 5\,{\rm yr}$, we find the reach of two LISA-like detectors as a dashed green line in Figure~\ref{fig:summary}.
For the frequency range, we use $f_{\rm min} = 10^{-5}\,{\rm Hz}$ and $f_{\rm max} =1 \,{\rm Hz}$.
We have included a galactic confusion noise, assuming $T=4\,{\rm yr}$ of operation~\cite{Cornish:2017vip}.

\subsubsection{\texorpdfstring{$\mu$}{}Ares} 
We also consider the $\mu$Ares proposal~\cite{Sesana:2019vho} for the ultralight dark matter search. 
The proposed concept is similar to LISA in the sense that it consists of three spacecraft, and it operates as a time-delay interferometer.
The configuration of satellites is heliocentric, while the arm-length is $L =400{\rm M}\,{\rm km}$. 
For the matched filter and cross-correlation analysis, we take the strain noise power spectrum in~\cite{Sesana:2019vho}, including astrophysical foreground. 
The upper limit on the dark matter density $\rho/\rho_0$ with matched filtering is shown as a purple line in Figure~\ref{fig:summary}. 

We also compare the ultralight dark matter-induced noise with the required noise level for the proposed mission concept. 
The proposal requires $S_{\rm acc}^{1/2}(f) = 10^{-15}\,{\rm m\,s^{-2}/\sqrt{Hz}}$, and  $S^{1/2}_N(f) = 50\,{\rm pm}/\sqrt{\rm Hz}$. 
The result is shown on the right panel of Figure~\ref{fig:ULDM_noise} for $m=10^{-12}\textrm{--} 10^{-15}\,{\rm eV}$. 
Note that the noise curve in this figure is obtained by using~\eqref{PS_X_TDI} with the proposal requirements, but without any astrophysical foregrounds. 
Note also that, for $f \lesssim \textrm{a few} \times 10^{-7}\,{\rm Hz}$, the gravity gradient noise from asteroids could dominate the total noise spectrum~\cite{Fedderke:2020yfy}. 

For the cross-correlation analysis, we specify the position of each spacecraft in the heliocentric ecliptic coordinate as
\begin{align}
\vec x_n = R \Big(\! \cos[\phi +  2 \pi (n-1)/ 3], \, \sin[\phi + 2\pi(n-1)/3] , \, 0 \Big)
\end{align}
where $n=1,2,3$ and $\phi$ is the orbital phase of the interferometer. 
We assume another of such constellations lying on the same plane (co-planar configuration), with the relative phase with respect to the first constellation by $\Delta \phi$. 
Using~\eqref{rho_cross}, we find the projected sensitivity of $\rho/\rho_0$ of the cross-correlation search in $\mu$Ares as a purple dashed line in Figure~\ref{fig:summary}.
We use $f_{\rm min} = 4\times 10^{-7}\,{\rm Hz}$, $f_{\rm max}= 10^{-1} \,{\rm Hz}$, and $\Delta \phi = \pi/2$.

\subsubsection{Big Bang Observatory}
The configuration of the Big Bang Observatory is similar to the LISA, except for a smaller arm-length, $L_{\rm BBO} =5\times 10^7\,{\rm m}$.
The instrument parameters are~\cite{Crowder:2005nr}
\bea
S_N^{1/2} &=& 1.4\times 10^{-17}\,{\rm m/\sqrt{Hz}}, 
\\
S_{\rm acc}^{1/2} &=& 3\times 10^{-17}\,{\rm m \,s^{-2}/\sqrt{Hz}} .
\eea
The cross-correlation search is identical to the one in LISA, except for this arm-length difference and different instrumental parameters.
The matched filter and cross-correlation search (using two co-planar constellations) yields the upper limit on $\rho/\rho_0$, represented by the (dashed) brown lines in Figure~\ref{fig:summary}. 
For the analysis, we use $f_{\rm min} = 10^{-4}\,{\rm Hz}$, $f_{\rm max} = 10^2\,{\rm Hz}$, and $\Delta \lambda = \pi$. 
We have included the same galactic confusion noise as in the case of LISA while assuming the other astrophysical foreground around $\sim {\cal O}(0.1)\,{\rm Hz}$ from neutron star binaries can be fully resolved~\cite{Yagi:2011wg}.

\section{Discussion}\label{sec:discussion}
We discuss now some of the ignored effects in the main discussion and some of the other implications of our results. 

In the main discussion, we have ignored the mean acceleration on the test masses due to ULDM; we have considered only its variance. 
The mean acceleration is nothing but the dynamical friction, and it is present as long as $\vec v_0\neq 0$ in the rest frame of the detector. 
One may directly compute the dynamical friction from the expression~\eqref{a}. 
The dynamical friction for the ultralight dark matter has already been obtained in~\cite{Bar-Or:2018pxz}, where it is given by
\bea
\langle a^i \rangle 
= \hat v_0^i \frac{8\pi G^2 \rho m_{\rm eff} \ln \Lambda}{\sigma^2} 
\left[ 
G\Big(\frac{v_0}{\sigma}\Big) + 
\frac{m_t}{2m_{\rm eff}} G\Big( \frac{v_0}{\sqrt{2} \sigma} \Big) \right]
\eea
Here $G(X) = 1/(2X^2) [ {\rm erf}(X) - (2X/\sqrt{\pi}) e^{-X^2} ]$, $m_t$ is the mass of the test body, and $\ln \Lambda$ is the Coulomb logarithm factor arising due to the long-range nature of gravitational force. 
For the halo dark matter with $\rho_0 =0.4\,{\rm GeV/cm^3}$ and $m = 10^{-22}\,{\rm eV}$, this dynamical friction force is causing $\Delta x \sim {\cal O}(1)\,{\rm meter}$ of drift of constellation over a year, which is much smaller than the expected unavoidable change of arm-length of LISA, $\Delta x \sim{\cal O}(10^4\textrm{--}10^{5})\,{\rm km}$, due to the orbital dynamics~\cite{LISA:2017pwj}. 
In addition, the dynamical friction force acts on each spacecraft in the same way, and therefore, we do not expect that it changes the arm length itself. 
For these reasons, we do not expect that the dynamical friction is relevant for the discussion of the ULDM-induced signals in the interferometers. 

We have found that some of the future gravitational wave interferometers could potentially probe $\rho / \rho_0 \sim {\cal O}(10^2)$ where $\rho_0 = 0.4\,{\rm GeV/cm^3}$ is the local dark matter density. 
Since we have projected the sensitivity of interferometers in units of $\rho_0$ and this local dark matter density is one of the key input parameters for all terrestrial dark matter detectors, it is worthwhile to discuss how this value $\rho_0 \sim {\cal O}(0.1)\,{\rm GeV/cm^3}$ is obtained. 
The current measurement of local dark matter is usually performed over a much larger spatial scale. 
For instance, even the very local measurements of the dark matter density select stellar tracers within ${\cal O}(10^2)$~pc around the solar system, and therefore, the inferred value of dark matter density is an average value over that spatial volume, $V \sim [{\cal O}(10^2)\,{\rm pc}]^3$~\cite{Read:2014qva, deSalas:2020hbh}. 
When it comes to the measurement within the solar system, the dark matter density is poorly constrained.
Measurement using the planetary ephemerides places an upper limit on the dark matter density in the solar system as $\rho/\rho_0 \lesssim 10^4$~\cite{Pitjev:2013sfa}, while the measurements of lunar laser ranging and LAGEOS geodetic satellite place an upper limit on the dark matter density between the Moon and the satellite's orbital radius as $\rho/\rho_0 \lesssim 10^{11}$~\cite{Adler:2008rq}. 

Furthermore, a large density fluctuation might just arise statistically in the ultralight dark matter halo. 
For the density of the field, $\rho_\phi$, its underlying distribution is given by the exponential distribution, $p(\rho_\phi) d\rho_\phi = [\exp(-\rho_\phi/\rho_0)/\rho_0 ] d\rho_\phi$ (see Appendix~\ref{App:large}).
Given that the correlation of density field exponentially drops over $L \gtrsim 1/m\sigma$~\cite{Bar-Or:2018pxz}, we may consider the density fluctuation $\rho_\phi$ over wavelength-sized patches as statistically independent random variables.
The probability of finding a patch with $\rho_\phi > c \rho_0$ is then given by
$$
P (\rho_\phi > c \rho_0) = \int_{c \rho_0}^\infty p(\rho_\phi) = e^{-c}. 
$$
On the other hand, the number of statistically independent patches over the volume $V$ is $N \sim V / \lambda^3$. 
For $m = 10^{-15}\eV$, the wavelength is about AU scale, and the number of patches over $V= (10\,{\rm kpc})^3$ volume is $N =V/\lambda^3= ( 10\,{\rm kpc}/{\rm AU} )^3 \sim 10^{28}$. 
For $ c = 60$, one finds $P(\rho_\phi > 60 \rho_0) = e^{-60} \simeq 10^{-26}$.
This would mean that there might be several hundreds of AU-sized patches within $10\,{\rm kpc}$ volume with $\rho \gtrsim 60\rho_0$.

We have only considered cross-correlation searches for space-borne interferometers.
The reason that it was not suitable for the ground-based interferometers, especially for the current advanced LIGO, was that the detector separation is too large for the ultralight dark matter signals to be correlated for the relevant frequency range of the detector.
However, in the initial LIGO phase, two detectors at Hanford, with $4\,$km and $2\,$km arms, were co-located and co-aligned, and therefore, it is in principle possible to use the result of two Hanford detectors in the initial LIGO phase for the cross-correlation search as demonstrated in~\cite{LIGOScientific:2014gqo, LIGOScientific:2014sej} for the stochastic gravitational waves.
Since the detectors are co-located, environmental noises are correlated, and this correlated noise must be carefully accounted for in the analysis.

For the low-frequency ULDM signals, we have considered the cross-correlation, which requires more than two detectors.
One may still be able to search for these low-frequency ULDM signals with a single detector with the help of a null data stream.
A null data stream is the output of the detector where the expected signal, either GWs or ULDM, is expected to be suppressed, allowing a way to calibrate the instrumental noises. 
For the time-delay interferometers, like LISA, a symmetrized Sagnac channel is shown to be insensitive to the gravitational wave signals at low frequencies, $f < f_* = (2\pi L)^{-1}$, which could, at least partially, provide a way to distinguish for stochastic gravitational wave backgrounds from other noise sources at such low frequencies~\cite{Tinto:2001ii, Hogan:2001jn, Romano:2016dpx}.
Similarly to the stochastic gravitational wave backgrounds, ultralight dark matter signal is strongly suppressed in the symmetrized Sagnac channel as $\propto (m \sigma L)^4$ for $m\sigma L < 1$, and therefore, using such null data stream might provide a way to search for ULDM at sufficiently small masses with the help of additional information such as annual modulations of the ULDM signal.

\section{Conclusion}\label{sec:conclusion}
We have investigated how the density fluctuations of ultralight dark matter interact with interferometers, designed for the detection of gravity waves. 
We provide a systematic way to compute the ULDM spectrum in any gravitational wave interferometers.
We show that the ultralight dark matter-induced noise is most significant when the arm-length is large, such as LISA or other proposals with an astronomical size of arm-length, and that, in all cases, we consider, the ultralight dark matter effects are subdominant compared to other noise sources, e.g. interferometric read-out noise and acceleration noise. 
Then we consider if such interferometers can be used to place an upper bound on the dark matter density in the solar system, and find that, under certain assumptions, $\rho/\rho_0 \sim{\cal O}(10^2)$ might be probed with future gravitational wave interferometers with its arm-length close to astronomical unit. 

We note that the current local dark matter density measurement, $\rho_0 =0.4\,{\rm GeV/cm^3}$, is performed on a much larger scale, e.g. $\gtrsim {\cal O}(100)\,{\rm pc}$, and that the constraints on dark matter abundance in the solar system from planetary ephemerides and laser ranging experiments are several orders of magnitude larger than $\rho_0$. 
Especially in the light of a situation where there is no direct and gravitational detection of dark matter in the solar system, these interferometric approaches will provide an interesting way to probe ultralight dark matter only through gravitational interaction. 
In addition, it has been recently proposed that the ultralight dark matter density in the solar system might be larger than the local dark matter density through certain capture processes~\cite{Banerjee:2019epw, Banerjee:2019xuy, Budker:2023sex}. 
Our interferometric search for ultralight dark matter in the solar system is expected to provide an interesting probe for such possibilities. 

Our analysis can be straightforwardly generalized to pulsar timing array (PTA) observations. Coherent ULDM signals in PTA observations were already explored by Khmelnitsky and Rubakov~\cite{Khmelnitsky:2013lxt}, whose results are reproduced in Appendix~\ref{app:pta} with the formalism discussed in this work. Low-frequency stochastic fluctuations of ULDM in the context of pulsar timing measurements, however, have not been explored. As the distance between the Earth and pulsars is on a kiloparsec scale, and the pulsar timing analyses involve a cross-correlation of timing residuals of pulsars with different sky locations, PTAs offer another interesting probe for stochastic ULDM density fluctuations. In future work, we will investigate how these stochastic ULDM signals affect pulsar timing measurements and if one can probe ULDM density within the solar system with PTA observations~\cite{KimMitridate}.

\acknowledgments
We would like to thank Juli{\' a}n Rey and Nicholas Rodd for useful conversations.
We especially thank Alessandro Lenoci for the initial collaboration.
We also thank Thomas Konstandin, Germano Nardini, and Mauro Pieroni for their useful comments and suggestions on the manuscript. 
This work is supported by the Deutsche Forschungsgemeinschaft under Germany’s Excellence Strategy - EXC 2121 Quantum Universe - 390833306.

\appendix

\section{Power spectrum}\label{App:PS}
We provide a detailed computation of the low-frequency part of the power spectrum $S_{\nabla\Psi}^{ij}(\omega, \vec L)$. 
To prepare an explicit computation, we derive the integral expression of the power spectrum \eqref{S_dP} in the non-relativistic limit:
\begin{align}
S_{\nabla \Psi}^{ij}(\omega,\vec L)
= & 32 \pi^3 G^2 
\int d^3 v_c \int d^3 v_d \,
f(\vec v_c + \vec v_d) 
f(\vec v_c - \vec v_d)
\label{S_a}
\\
\times & 
\Big[ 
v_c^i v_c^j 
\Big(
\delta(\omega - \omega_1 - \omega_2 ) e^{2 i m \vec v_c \cdot \vec L}
+ \delta(\omega + \omega_1 + \omega_2 ) e^{- 2 i m \vec v_c \cdot \vec L}
\Big)
\nonumber\\
& 
+ \frac{ v_d^i v_d^j}{v_d^4}
\Big( \delta(\omega - \omega_1 + \omega_2 ) 
e^{2 i m \vec v_d \cdot \vec L}
+ \delta(\omega + \omega_1 - \omega_2 ) 
e^{- 2 i m \vec v_d \cdot \vec L}
\Big)
\Big]
\nonumber
\end{align}
where we have introduced new integration variables $\vec v_c$ and $\vec v_d$, defined as
$$
\vec v_c = \frac{1}{2} (\vec v_1 + \vec v_2) , 
\qquad
\vec v_d = \frac{1}{2} (\vec v_1 - \vec v_2) .
$$
The two terms in the first line are related by the complex conjugate with $\omega \to - \omega$.
The two terms in the last line are the same under the exchange of the integration variable $\vec v_1 \leftrightarrow \vec v_2$. 
The first two terms represent the spectrum at $\omega = \pm 2m$, denoted by $A^{ij}(\omega, \vec L)$ in the main text, and the last two terms represent the low-frequency part, denoted by $B^{ij}(\omega, \vec L)$. 
Since the computation of $\omega = \pm 2m$ mode is straightforward, only the computation of $B^{ij}$ will be given below. 

The relevant part of the power spectrum is 
\begin{align}
S_{\nabla\Psi}^{ij}(\omega,\vec L)
= & 
\frac{32 \pi^3 G^2}{m}
\int d^3 v_c \int d^3 v_d \,
f(\vec v_c + \vec v_d) 
f(\vec v_c - \vec v_d)
\frac{ v_d^i v_d^j}{v_d^4}
\delta\Big(\vec v_c \cdot \vec v_d - \frac{\omega}{2m} \Big) 
e^{2 i m \vec v_d \cdot \vec L}
\end{align}
where we have used the symmetry of integrand under the exchange of integration variable $v_1\leftrightarrow v_2$ in \eqref{S_a} and take the non-relativistic limit for the $\delta$-function.\footnote{In the case of $\vec L=0$ and $\omega \to 0$ the above power spectrum is nothing but the second diffusion coefficient $D[\Delta v^i(T) \Delta v^j(T)] = \langle \Delta v^i(T) \Delta v^j(T)\rangle /T$ of the test mass with the change of velocity $\Delta v^i(T)$ over the time $T$. Compare the expression with (54) in Bar-Or et al~\cite{Bar-Or:2018pxz}.}
Using the normal distribution
$$
f(\vec v) = \frac{(\rho_0/m)}{(2\pi \sigma^2)^3} 
\exp\left[ - \frac{(\vec v - \vec v_0)^2}{2\sigma^2} \right],
$$
we find
\begin{align}
S_{\nabla\Psi}^{ij}(\omega,\vec L)
= & 
\frac{4 (G \rho_0)^2}{m^3 \sigma^6}
\int d^3 v_c \int d^3 v_d \,
\exp\left[
- \frac{(\vec v_c - \vec v_0)^2}{\sigma^2}
- \frac{v_d^2}{\sigma^2} \right]
\frac{ v_d^i v_d^j}{v_d^4}
\delta\Big(\vec v_c \cdot \vec v_d - \frac{\omega}{2m} \Big) 
e^{2 i m \vec v_d \cdot \vec L}
\nonumber\\
= & 
\frac{4 (G \rho_0)^2}{m^3\sigma^4}
\int d^3 v_c \int d^3 v_d \,
e^{- (\vec v_c - \vec v_0/\sigma )^2 }
e^{-v_d^2}
\frac{ v_d^i v_d^j}{v_d^4}
\delta\Big(\vec v_c \cdot \vec v_d - \frac{\bar \omega}{2} \Big) 
e^{2 i \vec v_d \cdot \vec L_\lambda}
\nonumber\\
= & 
- \frac{(G \rho_0)^2}{m^3 \sigma^4}
\frac{\partial^2}{\partial L_\lambda^i \partial L_\lambda^j}
\int d^3 v_c \int d^3 v_d \,
e^{- (\vec v_c - \vec v_0/\sigma)^2 }
e^{-v_d^2}
\frac{1}{v_d^4}
\delta\Big(\vec v_c \cdot \vec v_d - \frac{\bar \omega}{2} \Big) 
e^{2 i \vec v_d \cdot \vec L_\lambda}
\end{align}
where in the second line we have changed the integration variables $\vec v_{c,d} \to \vec v_{c,d} \sigma$, and defined $\vec L_\lambda = m \sigma \vec L$. 
In the last line, the vector $v_d^i$ in the integrand is replaced with the derivative with respect to $\vec L_\lambda$. 

To proceed further, we use the integral representation of $\delta$-function: $\delta(x) = \int (ds/2\pi) e^{-i x s}$. 
We then perform the $\vec v_c$ integral, and then the $\hat v_d$ integral. 
After some manipulation, we find
\begin{align}
S_{\nabla\Psi}^{ij}(\omega,\vec L)
= & 
- 4\pi^{3/2} \frac{(G \rho_0)^2}{m^3 \sigma^4}
\frac{\partial^2}{\partial L_\lambda^i \partial L_\lambda^j}
\int_{-\infty}^\infty ds \, e^{-i s \bar \omega} \sqrt{1+s^2}
\int_0^\infty \frac{e^{-y^2}}{y^2}
{\rm sinc} (2 y Y) 
\end{align}
where $\vec Y = (1+s^2)^{-1/2} (s \vec v_0/\sigma + \vec L_\lambda )$.
The $y$ integral is divergent.
The divergence can be regulated by ${\rm sinc}(2y Y) \to {\rm sinc}(2y Y) -1$. 
The additional $-1$ term would not contribute to the final result as it does not depend on $L_\lambda$. 
Using this regularization, we obtain
\begin{align}
S_{\nabla\Psi}^{ij}(\omega,\vec L)
= & 
(\bar a^2 \tau)
\left[ \frac{2}{\sqrt{\pi}}
\int_{-\infty}^\infty \frac{e^{-i s \bar\omega}}{\sqrt{1+s^2}}
\left(
\delta^{ij}
\frac{ {\rm erf} (Y) - G(Y) }{Y}
+ Y^i Y^i \frac{3 G(Y) - {\rm erf}(Y)}{Y^3} 
\right)
\right]
\nonumber\\
\equiv & (\bar a^2 \tau) B^{ij} (\omega, \vec L). 
\end{align}
This reproduces the result used in the main text. 

\section{Induced phase}\label{app:phase}
In this Appendix, we discuss the approximation for the phase \eqref{D_phi}.
We begin with
\bea
\Delta \phi (t) = 2 \omega_L ( \delta t + \delta L)
\eea
where the time delay $\delta t$ and the arm-length fluctuation $\delta L$ are
\begin{align}
\delta L & = 
+ \frac{1}{2} \hat x \cdot \Big[ 
\big( \delta \vec x_1( t- L) - \delta \vec x_0 (t-2L) \big)
+ \big( \delta \vec x_1( t- L) - \delta \vec x_0 (t) \big) \Big] , 
\label{dL}
\\
\delta t & = - \frac{1}{2}  \left[ \int^{t}_{t-L} dt' + \int^{t-L}_{t-2L} dt' \right] 
[ \Psi(t',\vec x(t')) + \Phi(t',\vec x(t'))  ] .
\label{dt}
\end{align}
In the Fourier space, each of them is
\begin{align}
\widetilde{\delta L} (\omega) &=
e^{i \omega L} \hat x \cdot 
\big[ \delta \vec x_1( \omega) 
-  \cos(\omega L) \delta \vec x_0 (\omega) \big]
=
\frac{ e^{i \omega L} }{ \omega^2 }
\int \frac{d^3k}{(2\pi)^3} e^{i \vec k \cdot \vec x_0}
\Big[ e^{i \vec k \cdot \hat n L} - \cos(\omega L)
\Big]
(i \vec k  \cdot \hat n)  \tilde \Phi
\\
\widetilde{\delta t} (\omega) & =
- e^{i \omega L}
\int \frac{d^3k}{(2\pi)^3} e^{i\vec k \cdot \vec x_0}
\left[
\frac{(e^{i \vec k \cdot \hat n L} -\cos\omega L) (i \vec k \cdot \hat n) + \omega \sin \omega L}{\omega^2 - (\vec k \cdot \hat n)^2} 
\right]
(\tilde \Phi + \tilde \Psi) 
\end{align}
where we have used $\delta \vec x_i(\omega) = -(1/\omega)^2 \int [d^3k/(2\pi)^3] e^{i \vec k \cdot \vec x_i} \tilde a(\omega, \vec k)$ and $\tilde a(k) = - i \vec k \tilde \Phi (k)$. 
Here $\hat n = (\vec x_1 - \vec x_0) / L$. 
The phase in the frequency space is
\begin{align}
\widetilde{\Delta\phi}(\omega) &=
2 e^{i \omega L} \omega_L
\int \frac{d^3k}{(2\pi)^3} e^{i \vec k \cdot \vec x_0}
\Big[
\frac{
\big( e^{i \vec k \cdot \hat n L} - \cos \omega L
\big)
(i \vec k  \cdot \hat n)}{\omega^2}  \tilde \Phi
- 
\frac{(e^{i \vec k \cdot \hat n L} -\cos\omega L) (i \vec k \cdot \hat n) + \omega \sin \omega L}{\omega^2 - (\vec k \cdot \hat n)^2} 
(\tilde \Phi + \tilde \Psi) 
\Big]. 
\end{align}

To simplify this expression, let us consider the power spectrum for the metric perturbation in the non-relativistic limit.
They are given as
\begin{align}
P_\Psi(k) 
&=
\frac{(4\pi G m)^2}{2 k^4}
(2\pi)^4 
\int d^3 v_1 d^3v_2
f_1 f_2
\Big[ 
v_c^4 \delta^{(4)}(k - p_1 - p_2)   + \delta^{(4)}(k - p_1 + p_2) 
\big)
\Big]
+ (k  \to - k)
\\
P_\Phi(k) 
&=
\frac{(4\pi G m)^2}{2k^4}
(2\pi)^4 
\int d^3 v_1 d^3 v_2 \, f_1f_2
\Big[
(v_c^2 + v_d^2 -3 (\hat v_c \cdot \vec v_d)^2 )^2 \delta^{(4)}(k - k_1 - k_2)
+ \delta^{(4)}(k - k_1 + k_2) 
\Big] + (k  \to - k)
\end{align}
where $f_i = f(\vec v_i)$ is the velocity distribution. 
In both cases, the $\omega =2m$ mode is velocity suppressed, while $\omega < m \sigma^2$ does not have such velocity suppression. 

Let us first consider $\omega < m \sigma^2$ modes. 
In this case, $\tilde \Phi = \tilde \Psi$ and $\omega / k \sim \sigma \ll 1$. 
Therefore, the time delay contribution is suppressed by $(\omega / k)^2\sim \sigma^2$ compared to the test mass acceleration.
Hence, the phase is approximated as
\bea
\widetilde{\Delta \phi}(\omega) 
=
e^{i \omega L} \frac{\omega_L}{\omega^2}
\int \frac{d^3k}{(2\pi)^3} e^{i \vec k \cdot \vec x_0}
(i \vec k \cdot \vec D) \tilde \Psi
\eea
with $\vec D(\hat n, L) = \hat n (e^{i \vec k \cdot \hat nL} -\cos \omega L)$.

Now consider $\omega =2m$. 
In this case, $\omega / k \sim 1/\sigma \gg 1$. 
The phase is approximated as
\begin{align}
\widetilde{\Delta\phi}(\omega) &\approx
- 2 e^{i \omega L} \frac{\omega_L }{\omega^2} 
\int \frac{d^3k}{(2\pi)^3} e^{i \vec k \cdot \vec x_0}
\Big[
 (e^{i \vec k \cdot \hat n L} -\cos\omega L) (i \vec k \cdot \hat n)\tilde \Psi
+  \omega \sin \omega L  \Big(1 + \frac{(\vec k \cdot \hat n)^2}{\omega^2} \Big) (  \tilde \Phi  + \tilde \Psi)
\Big]
\nonumber\\
&=
- 2 e^{i \omega L} \frac{\omega_L }{\omega^2} 
\int \frac{d^3k}{(2\pi)^3} e^{i \vec k \cdot \vec x_0}
\Big[
 (e^{i \vec k \cdot \hat n L} -\cos\omega L - i (\vec k \cdot \hat n L) \, {\rm sinc}\, \omega L ) (i \vec k \cdot \hat n)\tilde \Psi
+  {\rm sinc}\, \omega L  (\vec k \cdot \hat n)^2 L  \tilde \Phi 
\Big]
\end{align}
In the second line, we neglect the $\omega \sin(\omega L) (\tilde \Phi + \tilde \Psi)$ term.
Interferometers are always sensitive to the phase difference, and since this term does not have any dependence on the direction of light propagation, it cancels out in the final interferometer observables. 
For $\omega L \gtrsim 1$ and for $k L < (\omega L)^2 <1$, the terms proportional to $(\vec k \cdot \hat n)^2$ are negligible, and thus, the phase is approximated as
\begin{align}
\widetilde{\Delta\phi}(\omega) &\approx
- e^{i \omega L} \frac{\omega_L}{\omega^2}
\int \frac{d^3k}{(2\pi)^3} e^{i \vec k \cdot \vec x_0}
(i \vec k \cdot \vec D) \tilde \Psi
\label{app_high}
\end{align}
For $(\omega L)^2 < kL$, the dominant term is the one proportional to $\tilde \Phi$, so a more correct approximation would be obtained by replacing $\tilde \Psi \to - \tilde \Phi$ from \eqref{app_high}. 
Since $\tilde \Phi \sim \tilde \Psi$ up to an order one numerical factor, the above expression still provides a reasonable approximation even in this limit. 
In summary, the phase in both cases can be approximated as
\bea
\widetilde{\Delta \phi}(\omega) 
=
\pm e^{i \omega L} \frac{\omega_L}{\omega^2}
\int \frac{d^3k}{(2\pi)^3} e^{i \vec k \cdot \vec x_0}
(i \vec k \cdot \vec D) \tilde \Psi. 
\eea

\section{Density fluctuation}\label{App:large}
We compute the statistical properties of (mass) density fluctuation. 
In the non-relativistic limit, the density field becomes
\bea
\hat \rho_\phi
\approx
\sum_{i,j} \frac{m}{ V }
a_i^\dagger a_{j} e^{i ( k_i - k_j )\cdot x} ,
\eea
where we have ignored quantum fluctuations arising from the commutation relation. 
From the above expression with the quasi-probability distribution \eqref{quasi_prob}, moments of the density field can be computed as
\bea
\langle \rho_\phi^n \rangle
= n! \rho_0^n . 
\eea
They are identical to those random variables under the exponential distribution.
Therefore, we find
\bea
p(\rho_\phi) d \rho_\phi = \frac{e^{-\rho_\phi/\rho_0} }{\rho_0} d \rho_\phi
\eea
with $\rho_\phi \in [ 0 , \infty)$. 
The density fluctuation in the ultralight dark matter halo is larger than the particle dark matter halo since particle dark matter density follows the Poisson statistics.

\section{Coherent signals in pulsar timing array}\label{app:pta}
The pulsar timing array (PTA) measures the time-of-arrival of pulses from pulsars. 
Similar to the previous discussion, the time-of-arrival is affected by ULDM fluctuations through Shapiro time delay, Doppler effect, and Einstein delay. 
An arrival time of a pulse is given by
\begin{align}
\tau &\simeq \tau_{\rm em} + L + \delta\tau_1 + \delta \tau_2 + \delta\tau_3
\end{align}
where $\tau$ is the proper time of an observer at the observing time, $\tau_{\rm em}$ is the proper time at the pulsar at the emission time, $L$ is the nominal distance between the pulsar and the earth, and 
\begin{align}
\delta \tau_1(t) &= 
- \hat n \cdot \big[\delta \vec x_{\oplus}(t) -  \delta \vec x_{\rm psr}(t-L) \big],
\\
\delta \tau_2(t) &= - \int_{t-L}^t dt' \,  \big[ \Phi(t', \vec x(t')) + \Psi(t', \vec x(t')) \big],
\\
\delta \tau_3(t) &= 
\int^t  dt' \,
\Big[ \Phi(t', \vec x_\oplus ) -  \Phi(t' - L, \vec x_{\rm psr} ) ]. 
\end{align}
Here $\delta\tau_1$ is due to the change in distance between earth-pulsar system similar to \eqref{dL} (Doppler effect), $\delta \tau_2$ is the Shapiro time delay similar to \eqref{dt}, and $\delta\tau_3$ is due to the conversion between the proper time $\tau$ and the coordinate time $t$ (Einstein delay). 
Note that $\hat n$ is the unit vector pointing to the pulsar.

The next pulse arrives at $\tau'$,
\begin{align}
\tau' &\simeq \tau_{\rm em} + L + T
+ \delta\tau_1(t+T)
+ \delta\tau_2(t+T)
+ \delta\tau_3(t+T)
\end{align}
where $T$ is the rotational period of the pulsar 
The difference in time between the two arrival times is
\bea
\tau' - \tau  = T + \Delta T
\eea
where the time delay $\Delta T$ is 
\begin{align}
\Delta T &= 
\sum_{i=1}^3 
\Big[ \delta\tau_i(t+L) - \delta\tau_i(t) \Big] . 
\end{align}

The rotational period of pulsars used in PTA analysis is a few milliseconds, $T\sim{\cal O}(10^{-3})\,{\rm sec}$, while the coherent signal we are looking for has a frequency of a few nano Hz, $f \sim {\rm nHz}$. 
This suggests that the fluctuations due to ULDM do not significantly change over the rotational time scale $T$.
In the limit $f T \ll 1$, the time decay can be simplified as
\begin{align}
\frac{\Delta T}{T} (t)&= 
- \hat n \cdot \big[\delta \vec v_{\oplus}(t) -  \delta \vec v_{\rm psr}(t-L) \big]
- \Psi (t, \vec x_\oplus) + \Psi(t-L, \vec x_{\rm psr}) 
\nonumber\\
&
+ \int_{t-L}^{t} dt' \,  \hat n \cdot \nabla \Big[ \Phi(t',\vec x(t')) + \Psi(t',\vec x(t')) \Big].
\end{align}
This (approximate) time delay expression coincides with the redshift of the photon up to the sign~\cite{Khmelnitsky:2013lxt, Maggiore:2018sht}. 
Note that the Doppler shift was ignored in the work of Khmelnitsky and Rubakov~\cite{Khmelnitsky:2013lxt}.

The timing residual is computed by integrating the above time delay over time:
\bea
R(t) =\int_0^t dt' \frac{\Delta T}{T} (t'). 
\eea
For the coherent oscillation at $\omega = 2m$, the largest contribution to the timing residual arises from $- \Psi(t,\vec x_\oplus) + \Psi(t - L, \vec x_{\rm psr})$ part in the time delay. 
In particular, one finds
\bea
\langle R(t)^2 \rangle
\approx 
\int \frac{d^4k}{(2\pi)^4}
\frac{2}{\omega^4} P_\Psi(k)
\eea
where we have ignored fast-oscillating terms. 
The power spectrum is given by $P_\Psi(k) = [(4\pi G)^2/k^4] P_{\delta \rho}(k)$, where the density fluctuation power spectrum around $\omega \simeq 2m$ is
\begin{align}
 P_{\delta \rho}(k)
= \frac{2\pi^2 \rho^2}{m^4\sigma^5}
\left[
\Big(
\frac{v_k^4}{16} 
\Big( \frac{\bar v^2}{\sigma^2} - \frac{v_k^2}{4\sigma^2} \Big)^{1/2}
e^{-\bar v^2/\sigma^2}
+ (\omega \to - \omega) 
\Big)
+ \frac{\sigma}{v_k}
\exp\left( - \frac{v_k^2}{4\sigma^2} - \frac{\sigma^2|\bar\omega|^2}{v_k^2} \right)
\right]
\end{align}
The first term in the squared parenthesis represents the coherently oscillating mode at $\omega =2m$, while the second term represents the stochastic modes. 
This expression assumes $v_0=0$. 
Note also that $\bar v^2(\omega) = (\omega/m-2)$, $v_k = k / m$, and $\bar\omega = \omega/m \sigma^2$.
An explicit computation with the coherently oscillating modes reveals 
\bea
\langle R(t)^2 \rangle  \approx \frac{\pi^2 (G \rho)^2}{2m^6}
\eea
which agrees with the result in Ref.~\cite{Khmelnitsky:2013lxt}. 
While our main discussion focuses on computing the differential phase induced by ULDM in an interferometric setup, this exercise shows that the discussion presented in the main text is consistent with previous works.

\bibliographystyle{JHEP}
\bibliography{ref}

\providecommand{\href}[2]{#2}\begingroup\raggedright\begin{thebibliography}{10}

\bibitem{Peccei:1977hh}
R.D.~Peccei and H.R.~Quinn, \emph{{CP Conservation in the Presence of
  Instantons}}, \href{https://doi.org/10.1103/PhysRevLett.38.1440}{\emph{Phys.
  Rev. Lett.} {\bfseries 38} (1977) 1440}.

\bibitem{Weinberg:1977ma}
S.~Weinberg, \emph{{A New Light Boson?}},
  \href{https://doi.org/10.1103/PhysRevLett.40.223}{\emph{Phys. Rev. Lett.}
  {\bfseries 40} (1978) 223}.

\bibitem{Wilczek:1977pj}
F.~Wilczek, \emph{{Problem of Strong $P$ and $T$ Invariance in the Presence of
  Instantons}}, \href{https://doi.org/10.1103/PhysRevLett.40.279}{\emph{Phys.
  Rev. Lett.} {\bfseries 40} (1978) 279}.

\bibitem{Preskill:1982cy}
J.~Preskill, M.B.~Wise and F.~Wilczek, \emph{{Cosmology of the Invisible
  Axion}}, \href{https://doi.org/10.1016/0370-2693(83)90637-8}{\emph{Phys.
  Lett. B} {\bfseries 120} (1983) 127}.

\bibitem{Abbott:1982af}
L.F.~Abbott and P.~Sikivie, \emph{{A Cosmological Bound on the Invisible
  Axion}}, \href{https://doi.org/10.1016/0370-2693(83)90638-X}{\emph{Phys.
  Lett. B} {\bfseries 120} (1983) 133}.

\bibitem{Dine:1982ah}
M.~Dine and W.~Fischler, \emph{{The Not So Harmless Axion}},
  \href{https://doi.org/10.1016/0370-2693(83)90639-1}{\emph{Phys. Lett. B}
  {\bfseries 120} (1983) 137}.

\bibitem{Graham:2015cka}
P.W.~Graham, D.E.~Kaplan and S.~Rajendran, \emph{{Cosmological Relaxation of
  the Electroweak Scale}},
  \href{https://doi.org/10.1103/PhysRevLett.115.221801}{\emph{Phys. Rev. Lett.}
  {\bfseries 115} (2015) 221801}
  [\href{https://arxiv.org/abs/1504.07551}{{\ttfamily 1504.07551}}].

\bibitem{Arvanitaki:2016xds}
A.~Arvanitaki, S.~Dimopoulos, V.~Gorbenko, J.~Huang and K.~Van~Tilburg,
  \emph{{A small weak scale from a small cosmological constant}},
  \href{https://doi.org/10.1007/JHEP05(2017)071}{\emph{JHEP} {\bfseries 05}
  (2017) 071} [\href{https://arxiv.org/abs/1609.06320}{{\ttfamily
  1609.06320}}].

\bibitem{Banerjee:2018xmn}
A.~Banerjee, H.~Kim and G.~Perez, \emph{{Coherent relaxion dark matter}},
  \href{https://doi.org/10.1103/PhysRevD.100.115026}{\emph{Phys. Rev. D}
  {\bfseries 100} (2019) 115026}
  [\href{https://arxiv.org/abs/1810.01889}{{\ttfamily 1810.01889}}].

\bibitem{Banerjee:2020kww}
A.~Banerjee, H.~Kim, O.~Matsedonskyi, G.~Perez and M.S.~Safronova,
  \emph{{Probing the Relaxed Relaxion at the Luminosity and Precision
  Frontiers}}, \href{https://doi.org/10.1007/JHEP07(2020)153}{\emph{JHEP}
  {\bfseries 07} (2020) 153}
  [\href{https://arxiv.org/abs/2004.02899}{{\ttfamily 2004.02899}}].

\bibitem{Arkani-Hamed:2020yna}
N.~Arkani-Hamed, R.T.~D'Agnolo and H.D.~Kim, \emph{{Weak scale as a trigger}},
  \href{https://doi.org/10.1103/PhysRevD.104.095014}{\emph{Phys. Rev. D}
  {\bfseries 104} (2021) 095014}
  [\href{https://arxiv.org/abs/2012.04652}{{\ttfamily 2012.04652}}].

\bibitem{TitoDAgnolo:2021nhd}
R.~Tito~D'Agnolo and D.~Teresi, \emph{{Sliding Naturalness: New Solution to the
  Strong-$CP$ and Electroweak-Hierarchy Problems}},
  \href{https://doi.org/10.1103/PhysRevLett.128.021803}{\emph{Phys. Rev. Lett.}
  {\bfseries 128} (2022) 021803}
  [\href{https://arxiv.org/abs/2106.04591}{{\ttfamily 2106.04591}}].

\bibitem{TitoDAgnolo:2021pjo}
R.~Tito~D'Agnolo and D.~Teresi, \emph{{Sliding naturalness: cosmological
  selection of the weak scale}},
  \href{https://doi.org/10.1007/JHEP02(2022)023}{\emph{JHEP} {\bfseries 02}
  (2022) 023} [\href{https://arxiv.org/abs/2109.13249}{{\ttfamily
  2109.13249}}].

\bibitem{Chatrchyan:2022dpy}
A.~Chatrchyan and G.~Servant, \emph{{Relaxion Dark Matter from Stochastic
  Misalignment}},  \href{https://arxiv.org/abs/2211.15694}{{\ttfamily
  2211.15694}}.

\bibitem{Hu:2000ke}
W.~Hu, R.~Barkana and A.~Gruzinov, \emph{{Cold and fuzzy dark matter}},
  \href{https://doi.org/10.1103/PhysRevLett.85.1158}{\emph{Phys. Rev. Lett.}
  {\bfseries 85} (2000) 1158}
  [\href{https://arxiv.org/abs/astro-ph/0003365}{{\ttfamily
  astro-ph/0003365}}].

\bibitem{Hui:2021tkt}
L.~Hui, \emph{{Wave Dark Matter}},
  \href{https://doi.org/10.1146/annurev-astro-120920-010024}{\emph{Ann. Rev.
  Astron. Astrophys.} {\bfseries 59} (2021) 247}
  [\href{https://arxiv.org/abs/2101.11735}{{\ttfamily 2101.11735}}].

\bibitem{Schive:2014dra}
H.-Y.~Schive, T.~Chiueh and T.~Broadhurst, \emph{{Cosmic Structure as the
  Quantum Interference of a Coherent Dark Wave}},
  \href{https://doi.org/10.1038/nphys2996}{\emph{Nature Phys.} {\bfseries 10}
  (2014) 496} [\href{https://arxiv.org/abs/1406.6586}{{\ttfamily 1406.6586}}].

\bibitem{Schive:2014hza}
H.-Y.~Schive, M.-H.~Liao, T.-P.~Woo, S.-K.~Wong, T.~Chiueh, T.~Broadhurst
  et~al., \emph{{Understanding the Core-Halo Relation of Quantum Wave Dark
  Matter from 3D Simulations}},
  \href{https://doi.org/10.1103/PhysRevLett.113.261302}{\emph{Phys. Rev. Lett.}
  {\bfseries 113} (2014) 261302}
  [\href{https://arxiv.org/abs/1407.7762}{{\ttfamily 1407.7762}}].

\bibitem{Hui:2016ltb}
L.~Hui, J.P.~Ostriker, S.~Tremaine and E.~Witten, \emph{{Ultralight scalars as
  cosmological dark matter}},
  \href{https://doi.org/10.1103/PhysRevD.95.043541}{\emph{Phys. Rev. D}
  {\bfseries 95} (2017) 043541}
  [\href{https://arxiv.org/abs/1610.08297}{{\ttfamily 1610.08297}}].

\bibitem{Seto:2004zu}
N.~Seto and A.~Cooray, \emph{{Search for small-mass black hole dark matter with
  space-based gravitational wave detectors}},
  \href{https://doi.org/10.1103/PhysRevD.70.063512}{\emph{Phys. Rev. D}
  {\bfseries 70} (2004) 063512}
  [\href{https://arxiv.org/abs/astro-ph/0405216}{{\ttfamily
  astro-ph/0405216}}].

\bibitem{Adams:2004pk}
A.W.~Adams and J.S.~Bloom, \emph{{Direct detection of dark matter with
  space-based laser interferometers}},
  \href{https://arxiv.org/abs/astro-ph/0405266}{{\ttfamily astro-ph/0405266}}.

\bibitem{Hall:2016usm}
E.D.~Hall, R.X.~Adhikari, V.V.~Frolov, H.~M\"uller, M.~Pospelov and
  R.X.~Adhikari, \emph{{Laser Interferometers as Dark Matter Detectors}},
  \href{https://doi.org/10.1103/PhysRevD.98.083019}{\emph{Phys. Rev. D}
  {\bfseries 98} (2018) 083019}
  [\href{https://arxiv.org/abs/1605.01103}{{\ttfamily 1605.01103}}].

\bibitem{Jaeckel:2020mqa}
J.~Jaeckel, S.~Schenk and M.~Spannowsky, \emph{{Probing dark matter clumps,
  strings and domain walls with gravitational wave detectors}},
  \href{https://doi.org/10.1140/epjc/s10052-021-09604-9}{\emph{Eur. Phys. J. C}
  {\bfseries 81} (2021) 828}
  [\href{https://arxiv.org/abs/2004.13724}{{\ttfamily 2004.13724}}].

\bibitem{Baum:2022duc}
S.~Baum, M.A.~Fedderke and P.W.~Graham, \emph{{Searching for dark clumps with
  gravitational-wave detectors}},
  \href{https://doi.org/10.1103/PhysRevD.106.063015}{\emph{Phys. Rev. D}
  {\bfseries 106} (2022) 063015}
  [\href{https://arxiv.org/abs/2206.14832}{{\ttfamily 2206.14832}}].

\bibitem{Sesana:2019vho}
A.~Sesana et~al., \emph{{Unveiling the gravitational universe at $\mu$-Hz
  frequencies}}, \href{https://doi.org/10.1007/s10686-021-09709-9}{\emph{Exper.
  Astron.} {\bfseries 51} (2021) 1333}
  [\href{https://arxiv.org/abs/1908.11391}{{\ttfamily 1908.11391}}].

\bibitem{Fedderke:2021kuy}
M.A.~Fedderke, P.W.~Graham and S.~Rajendran, \emph{{Asteroids for
  \ensuremath{\mu}Hz gravitational-wave detection}},
  \href{https://doi.org/10.1103/PhysRevD.105.103018}{\emph{Phys. Rev. D}
  {\bfseries 105} (2022) 103018}
  [\href{https://arxiv.org/abs/2112.11431}{{\ttfamily 2112.11431}}].

\bibitem{Fedderke:2020yfy}
M.A.~Fedderke, P.W.~Graham and S.~Rajendran, \emph{{Gravity Gradient Noise from
  Asteroids}}, \href{https://doi.org/10.1103/PhysRevD.103.103017}{\emph{Phys.
  Rev. D} {\bfseries 103} (2021) 103017}
  [\href{https://arxiv.org/abs/2011.13833}{{\ttfamily 2011.13833}}].

\bibitem{Read:2014qva}
J.I.~Read, \emph{{The Local Dark Matter Density}},
  \href{https://doi.org/10.1088/0954-3899/41/6/063101}{\emph{J. Phys. G}
  {\bfseries 41} (2014) 063101}
  [\href{https://arxiv.org/abs/1404.1938}{{\ttfamily 1404.1938}}].

\bibitem{deSalas:2020hbh}
P.F.~de~Salas and A.~Widmark, \emph{{Dark matter local density determination:
  recent observations and future prospects}},
  \href{https://doi.org/10.1088/1361-6633/ac24e7}{\emph{Rept. Prog. Phys.}
  {\bfseries 84} (2021) 104901}
  [\href{https://arxiv.org/abs/2012.11477}{{\ttfamily 2012.11477}}].

\bibitem{Cai:2023ywp}
R.-G.~Cai, Z.-K.~Guo, B.~Hu, C.~Liu, Y.~Lu, W.-T.~Ni et~al., \emph{{On networks
  of space-based gravitational-wave detectors}},
  \href{https://arxiv.org/abs/2305.04551}{{\ttfamily 2305.04551}}.

\bibitem{Kim:2021yyo}
H.~Kim and A.~Lenoci, \emph{{Gravitational focusing of wave dark matter}},
  \href{https://doi.org/10.1103/PhysRevD.105.063032}{\emph{Phys. Rev. D}
  {\bfseries 105} (2022) 063032}
  [\href{https://arxiv.org/abs/2112.05718}{{\ttfamily 2112.05718}}].

\bibitem{Derevianko:2016vpm}
A.~Derevianko, \emph{{Detecting dark-matter waves with a network of
  precision-measurement tools}},
  \href{https://doi.org/10.1103/PhysRevA.97.042506}{\emph{Phys. Rev. A}
  {\bfseries 97} (2018) 042506}
  [\href{https://arxiv.org/abs/1605.09717}{{\ttfamily 1605.09717}}].

\bibitem{Foster:2017hbq}
J.W.~Foster, N.L.~Rodd and B.R.~Safdi, \emph{{Revealing the Dark Matter Halo
  with Axion Direct Detection}},
  \href{https://doi.org/10.1103/PhysRevD.97.123006}{\emph{Phys. Rev. D}
  {\bfseries 97} (2018) 123006}
  [\href{https://arxiv.org/abs/1711.10489}{{\ttfamily 1711.10489}}].

\bibitem{Centers:2019dyn}
G.P.~Centers et~al., \emph{{Stochastic fluctuations of bosonic dark matter}},
  \href{https://doi.org/10.1038/s41467-021-27632-7}{\emph{Nature Commun.}
  {\bfseries 12} (2021) 7321}
  [\href{https://arxiv.org/abs/1905.13650}{{\ttfamily 1905.13650}}].

\bibitem{Otto:2015erp}
M.~Otto, \emph{{Time-Delay Interferometry Simulations for the Laser
  Interferometer Space Antenna}}, Ph.D. thesis, Leibniz U., Hannover, 2015.
\newblock 10.15488/8545.

\bibitem{Tinto:2020fcc}
M.~Tinto and S.V.~Dhurandhar, \emph{{Time-delay interferometry}},
  \href{https://doi.org/10.1007/s41114-020-00029-6}{\emph{Living Rev. Rel.}
  {\bfseries 24} (2021) 1}.

\bibitem{1999PhRvD..59j2003T}
M.~{Tinto} and J.W.~{Armstrong}, \emph{{Cancellation of laser noise in an
  unequal-arm interferometer detector of gravitational radiation}},
  \href{https://doi.org/10.1103/PhysRevD.59.102003}{\emph{\prd} {\bfseries 59}
  (1999) 102003}.

\bibitem{1999ApJ...527..814A}
J.W.~{Armstrong}, F.B.~{Estabrook} and M.~{Tinto}, \emph{{Time-Delay
  Interferometry for Space-based Gravitational Wave Searches}},
  \href{https://doi.org/10.1086/308110}{\emph{\apj} {\bfseries 527} (1999)
  814}.

\bibitem{Babak:2021mhe}
S.~Babak, A.~Petiteau and M.~Hewitson, \emph{{LISA Sensitivity and SNR
  Calculations}},  \href{https://arxiv.org/abs/2108.01167}{{\ttfamily
  2108.01167}}.

\bibitem{Maggiore:2007ulw}
M.~Maggiore, \emph{{Gravitational Waves. Vol. 1: Theory and Experiments}},
  Oxford Master Series in Physics, Oxford University Press (2007).

\bibitem{BBO}
S.~Phinney~et al, \emph{{The Big Bang Observer: Direct Detection of
  Gravitational Waves from the Nirth of the Universe to the Presen}},
  {\emph{NASA Mission Concept Study} (2004) }.

\bibitem{aLIGO:2020wna}
{\scshape aLIGO} collaboration, \emph{{Sensitivity and performance of the
  Advanced LIGO detectors in the third observing run}},
  \href{https://doi.org/10.1103/PhysRevD.102.062003}{\emph{Phys. Rev. D}
  {\bfseries 102} (2020) 062003}
  [\href{https://arxiv.org/abs/2008.01301}{{\ttfamily 2008.01301}}].

\bibitem{Hild:2010id}
S.~Hild et~al., \emph{{Sensitivity Studies for Third-Generation Gravitational
  Wave Observatories}},
  \href{https://doi.org/10.1088/0264-9381/28/9/094013}{\emph{Class. Quant.
  Grav.} {\bfseries 28} (2011) 094013}
  [\href{https://arxiv.org/abs/1012.0908}{{\ttfamily 1012.0908}}].

\bibitem{LISA:2017pwj}
{\scshape LISA} collaboration, \emph{{Laser Interferometer Space Antenna}},
  \href{https://arxiv.org/abs/1702.00786}{{\ttfamily 1702.00786}}.

\bibitem{Armano:2016bkm}
M.~Armano et~al., \emph{{Sub-Femto- g Free Fall for Space-Based Gravitational
  Wave Observatories: LISA Pathfinder Results}},
  \href{https://doi.org/10.1103/PhysRevLett.116.231101}{\emph{Phys. Rev. Lett.}
  {\bfseries 116} (2016) 231101}.

\bibitem{Armano:2018kix}
M.~Armano et~al., \emph{{Beyond the Required LISA Free-Fall Performance: New
  LISA Pathfinder Results down to 20 $\mu$Hz}},
  \href{https://doi.org/10.1103/PhysRevLett.120.061101}{\emph{Phys. Rev. Lett.}
  {\bfseries 120} (2018) 061101}.

\bibitem{Robson:2018ifk}
T.~Robson, N.J.~Cornish and C.~Liu, \emph{{The construction and use of LISA
  sensitivity curves}},
  \href{https://doi.org/10.1088/1361-6382/ab1101}{\emph{Class. Quant. Grav.}
  {\bfseries 36} (2019) 105011}
  [\href{https://arxiv.org/abs/1803.01944}{{\ttfamily 1803.01944}}].

\bibitem{Cornish:2001bb}
N.J.~Cornish, \emph{{Detecting a stochastic gravitational wave background with
  the Laser Interferometer Space Antenna}},
  \href{https://doi.org/10.1103/PhysRevD.65.022004}{\emph{Phys. Rev. D}
  {\bfseries 65} (2002) 022004}
  [\href{https://arxiv.org/abs/gr-qc/0106058}{{\ttfamily gr-qc/0106058}}].

\bibitem{Cornish:2017vip}
N.~Cornish and T.~Robson, \emph{{Galactic binary science with the new LISA
  design}}, \href{https://doi.org/10.1088/1742-6596/840/1/012024}{\emph{J.
  Phys. Conf. Ser.} {\bfseries 840} (2017) 012024}
  [\href{https://arxiv.org/abs/1703.09858}{{\ttfamily 1703.09858}}].

\bibitem{Crowder:2005nr}
J.~Crowder and N.J.~Cornish, \emph{{Beyond LISA: Exploring future gravitational
  wave missions}},
  \href{https://doi.org/10.1103/PhysRevD.72.083005}{\emph{Phys. Rev. D}
  {\bfseries 72} (2005) 083005}
  [\href{https://arxiv.org/abs/gr-qc/0506015}{{\ttfamily gr-qc/0506015}}].

\bibitem{Yagi:2011wg}
K.~Yagi and N.~Seto, \emph{{Detector configuration of DECIGO/BBO and
  identification of cosmological neutron-star binaries}},
  \href{https://doi.org/10.1103/PhysRevD.83.044011}{\emph{Phys. Rev. D}
  {\bfseries 83} (2011) 044011}
  [\href{https://arxiv.org/abs/1101.3940}{{\ttfamily 1101.3940}}].

\bibitem{Bar-Or:2018pxz}
B.~Bar-Or, J.-B.~Fouvry and S.~Tremaine, \emph{{Relaxation in a Fuzzy Dark
  Matter Halo}},
  \href{https://doi.org/10.3847/1538-4357/aaf28c}{\emph{Astrophys. J.}
  {\bfseries 871} (2019) 28}
  [\href{https://arxiv.org/abs/1809.07673}{{\ttfamily 1809.07673}}].

\bibitem{Pitjev:2013sfa}
N.P.~Pitjev and E.V.~Pitjeva, \emph{{Constraints on dark matter in the solar
  system}}, \href{https://doi.org/10.1134/S1063773713020060}{\emph{Astron.
  Lett.} {\bfseries 39} (2013) 141}
  [\href{https://arxiv.org/abs/1306.5534}{{\ttfamily 1306.5534}}].

\bibitem{Adler:2008rq}
S.L.~Adler, \emph{{Placing direct limits on the mass of earth-bound dark
  matter}}, \href{https://doi.org/10.1088/1751-8113/41/41/412002}{\emph{J.
  Phys. A} {\bfseries 41} (2008) 412002}
  [\href{https://arxiv.org/abs/0808.0899}{{\ttfamily 0808.0899}}].

\bibitem{LIGOScientific:2014gqo}
{\scshape LIGO Scientific, VIRGO} collaboration, \emph{{Improved Upper Limits
  on the Stochastic Gravitational-Wave Background from 2009\textendash{}2010
  LIGO and Virgo Data}},
  \href{https://doi.org/10.1103/PhysRevLett.113.231101}{\emph{Phys. Rev. Lett.}
  {\bfseries 113} (2014) 231101}
  [\href{https://arxiv.org/abs/1406.4556}{{\ttfamily 1406.4556}}].

\bibitem{LIGOScientific:2014sej}
{\scshape LIGO Scientific, VIRGO} collaboration, \emph{{Searching for
  stochastic gravitational waves using data from the two colocated LIGO Hanford
  detectors}}, \href{https://doi.org/10.1103/PhysRevD.91.022003}{\emph{Phys.
  Rev. D} {\bfseries 91} (2015) 022003}
  [\href{https://arxiv.org/abs/1410.6211}{{\ttfamily 1410.6211}}].

\bibitem{Tinto:2001ii}
M.~Tinto, J.W.~Armstrong and F.B.~Estabrook, \emph{{Discriminating a
  gravitational wave background from instrumental noise in the LISA detector}},
  \href{https://doi.org/10.1103/PhysRevD.63.021101}{\emph{Phys. Rev. D}
  {\bfseries 63} (2001) 021101}.

\bibitem{Hogan:2001jn}
C.J.~Hogan and P.L.~Bender, \emph{{Estimating stochastic gravitational wave
  backgrounds with Sagnac calibration}},
  \href{https://doi.org/10.1103/PhysRevD.64.062002}{\emph{Phys. Rev. D}
  {\bfseries 64} (2001) 062002}
  [\href{https://arxiv.org/abs/astro-ph/0104266}{{\ttfamily
  astro-ph/0104266}}].

\bibitem{Romano:2016dpx}
J.D.~Romano and N.J.~Cornish, \emph{{Detection methods for stochastic
  gravitational-wave backgrounds: a unified treatment}},
  \href{https://doi.org/10.1007/s41114-017-0004-1}{\emph{Living Rev. Rel.}
  {\bfseries 20} (2017) 2} [\href{https://arxiv.org/abs/1608.06889}{{\ttfamily
  1608.06889}}].

\bibitem{Banerjee:2019epw}
A.~Banerjee, D.~Budker, J.~Eby, H.~Kim and G.~Perez, \emph{{Relaxion Stars and
  their detection via Atomic Physics}},
  \href{https://doi.org/10.1038/s42005-019-0260-3}{\emph{Commun. Phys.}
  {\bfseries 3} (2020) 1} [\href{https://arxiv.org/abs/1902.08212}{{\ttfamily
  1902.08212}}].

\bibitem{Banerjee:2019xuy}
A.~Banerjee, D.~Budker, J.~Eby, V.V.~Flambaum, H.~Kim, O.~Matsedonskyi et~al.,
  \emph{{Searching for Earth/Solar Axion Halos}},
  \href{https://doi.org/10.1007/JHEP09(2020)004}{\emph{JHEP} {\bfseries 09}
  (2020) 004} [\href{https://arxiv.org/abs/1912.04295}{{\ttfamily
  1912.04295}}].

\bibitem{Budker:2023sex}
D.~Budker, J.~Eby, M.~Gorghetto, M.~Jiang and G.~Perez, \emph{{A Generic
  Formation Mechanism of Ultralight Dark Matter Solar Halos}},
  \href{https://arxiv.org/abs/2306.12477}{{\ttfamily 2306.12477}}.

\bibitem{Khmelnitsky:2013lxt}
A.~Khmelnitsky and V.~Rubakov, \emph{{Pulsar timing signal from ultralight
  scalar dark matter}},
  \href{https://doi.org/10.1088/1475-7516/2014/02/019}{\emph{JCAP} {\bfseries
  02} (2014) 019} [\href{https://arxiv.org/abs/1309.5888}{{\ttfamily
  1309.5888}}].

\bibitem{KimMitridate}
H.~Kim and A.~Mitridate, \emph{{to appear}}, .

\bibitem{Maggiore:2018sht}
M.~Maggiore, \emph{{Gravitational Waves. Vol. 2: Astrophysics and Cosmology}},
  Oxford University Press (3, 2018).

\end{thebibliography}\endgroup
\end{document}